\documentclass[prb,amsmath,amssymb,superscriptaddress,twocolumn]{revtex4}

\usepackage{amsmath}
\usepackage{graphicx}
\usepackage{color}
\usepackage{float}
\usepackage{braket}
\usepackage{amssymb}
\usepackage{bbold}
\DeclareMathOperator{\ud}{\upharpoonleft\!\downharpoonright}

\begin{document}
\title{Ferromagnetism in Quantum Dot Plaquettes}
\author{Donovan Buterakos}
\author{Sankar Das Sarma}
\affiliation{Condensed Matter Theory Center and Joint Quantum Institute, Department of Physics, University of Maryland, College Park, Maryland 20742-4111 USA}
\date{\today}

\begin{abstract}
	Following the recent claimed obsevation of Nagaoka ferromagnetism in finite size quantum dot plaquettes,\cite{DehollainARXIV2019} a general theoretical analysis is warranted in order to ascertain in rather generic terms which arrangements of a small number of quantum dots can produce saturated ferromagnetic ground states and under which constraints on interaction and inter-dot tunneling in the plaquette.  This is particularly necessary since Nagaoka ferromagnetism is fragile and arises only under rather special conditions. We test the robustness of ground state ferromagnetism in the presence of a long-range Coulomb interaction and long-range as well as short-range interdot hopping by modeling a wide range of different plaquette geometries accessible by arranging a few ($\sim$4) quantum dots in a controlled manner. We find that ferromagnetism is robust to the presence of long range Coulomb interactions, and we develop conditions constraining the tunneling strength such that the ground state is ferromagnetic.  Additionally, we predict the presence of a partially spin-polarized ferromagnetic state for 4 electrons in a Y-shaped 4-quantum dot plaquette. Finally, we consider 4 electrons in a ring of 5 dots. This does not satisfy the Nagaoka condition, however, we show that the ground state is spin one for strong, but not infinite, onsite interaction. Thus, even though Nagaoka's theorem does not apply, the ground state for the finite system with one hole in a ring of 5 dots is partially ferromagnetic. We provide detailed fully analytical results for the existence or not of ferromagnetic ground states in several quantum dot geometries which can be studied in currently available coupled quantum dot systems.
\end{abstract}

\maketitle

\section{Introduction}

John Hubbard introduced the celebrated Hubbard model \cite{HubbardPRS1963} as a minimal model to study ferromagnetism in narrow band itinerant electron systems such as Fe, Ni, and Co.  The hope was that the minimal Hubbard model, with just one dimensionless interaction parameter $U/t$ where $U$ is the on-site interaction (arising from Coulomb repulsion) between two electrons with unlike spins and $t$ is the nearest-neighbor tunneling associated with kinetic energy, would make the difficult problem of itinerant electron metallic ferromagnetism tractable and perhaps even exactly solvable.  This early hope of the Hubbard model leading perhaps to an understanding of narrow band metallic ferromagnetism was echoed in other early publications also.\cite{KanamoriPTP1963, GutzwillerPRL1963}  After almost 60 years of extensive research, we still do not have a general solution to the Hubbard model (except under very restricted conditions, e.g., one dimensional, 1D, systems) and the Hubbard model has become the archetype underlying the whole subject of strongly correlated materials.  In fact, large teams of computational physicists work on large computers with the single goal of trying to understand numerically the implications of the Hubbard model in various situations, and no clear signatures for ferromagnetic ground states in the Hubbard model have emerged from these extensive numerical calculations.\cite{LeBlancPRX2015}  Perhaps the most ironic aspect of the Hubbard model is that it is now universally accepted to be an excellent model to study antiferromagnetism, local moment formation, and Mott metal-insulator transition in narrow band lattice systems rather than as a model for metallic ferromagnetism as Hubbard originally dreamed of. Any ferromagnetism arising within the Hubbard model is fragile and is certainly limited to very narrow parameter ranges (i.e. band filling and the interaction strength $U/t$), and it is entirely possible that generic 2D and 3D ferromagnetic systems cannot be described by the Hubbard model at all.

One important early result in this context is the concept of Nagaoka ferromagnetism\cite{NagaokaPR1966} which arises naturally in the 2D Hubbard model on square (and other bipartite) lattices under rather nongeneric and highly restrictive conditions (see, e.g. Refs. \onlinecite{TasakiPTP1998}, \onlinecite{BobrowPRB2018}, and references therein).  This is an exact result which asserts that the 2D Hubbard model doped by precisely one hole (i.e. one missing electron) away from the half-filling has full ferromagnetism of the whole system in the thermodynamic limit provided $U$ is infinite.  Since the half-filled 2D Hubbard model is surely not a ferromagnet at any interaction strength, the Nagaoka theorem appears pretty amazing in the sense that removing just one electron from the system drives the whole ground state completely ferromagnetic.  The theorem derives from the kinetic constraint on the motion of a hole in the half-filled system in the infinite $U$ limit, leading to the lowest energy state being the state of all the electrons becoming spin-polarized in order to minimize the kinetic energy in the strongly interacting limit (where double occupancy is not allowed).  While Nagaoka ferromagnetism is of some theoretical significance because it is an exact result, it is of no consequence for any experimental situation since creating precisely one hole in a thermodynamic system is obviously an impossible constraint (and the infinite interaction limit is unphysical as well).  The very fragile nature of the proof underlying this theorem does not allow its generalization to a dilute density of holes around half-filling, and Nagaoka ferromagnetism in its original form\cite{NagaokaPR1966} is unlikely to be observable experimentally in spite of its theoretical validity.

The question we address in the current work is the relevance of Nagaoka ferromagnetism in small finite 2D systems, which can be constructed by using semiconductor quantum dots with a few electrons in it.  In such a system, with $N$ electrons in $M$ dots, the effective finite-size Nagaoka situation is easily achieved by tuning the system to having $N=M-1$, assuming each dot to have one effective orbital energy level with two spin states.  Such a scenario was recently achieved experimentally in Ref. \onlinecite{DehollainARXIV2019}, and signatures for ferromagnetism were observed.  Our goal in the current work is to ask a general theoretical question on the existence or not of ferromagnetic ground states in small 2D plaquettes made of tunable semiconductor quantum dots: What experimentally accessible arrangements of a few coupled quantum dots ($\sim$4) with a few electrons would manifest stable ferromagnetic ground states?  It turns out that this question can be answered analytically for several interesting quantum dot structures which are currently experimentally viable because of recent advances in control, engineering, and fabrication of coupled semiconductor quantum dots in the context of developing spin qubits.\cite{DehollainARXIV2019,HensgensNAT2017,VanDiepenAPL2018,EeninkARXIV2019,MillsNC2019,MillsARXIV2019}

It was pointed out 25 years ago\cite{StaffordPRL1994,KotlyarPRB1998,Kotlyar2PRB1998} that semiconductor quantum dot arrays may be capable of simulating the Hubbard model in finite solid state systems searching for Mott transition and related strong correlation phenomena.  Advances in materials growth and nanofabrication techniques finally made this idea practical in laboratory settings only in 2017 when Mott physics in the form of the predicted collective Coulomb blockade\cite{StaffordPRL1994} was observed in a small linear array of coupled GaAs quantum dots emulating the Hubbard model.\cite{HensgensNAT2017}  There has been rapid recent development in controlling small coupled quantum dot arrays in several laboratories\cite{DehollainARXIV2019,HensgensNAT2017,VanDiepenAPL2018,EeninkARXIV2019,MillsNC2019,MillsARXIV2019}, and experimentalists can now study up to 4-8 dots with variable numbers of electrons per dot along with precise control of coherent electron tunneling between the dots. Our work, although purely theoretical, is inspired by these developments in the precise experimental control over small systems of coupled quantum dots.  In particular, the recent experimental work from Delft\cite{DehollainARXIV2019} reporting the observation of Nagaoka ferromagnetism in a 2D square array of quantum dots has directly motivated our work although our emphasis is on the generality of the possible emergence of Nagaoka-type ferromagnetism in quantum dot arrays, not describing the observations in Ref. \onlinecite{DehollainARXIV2019} which require a detailed numerical approach.\cite{WangARXIV2019}

Electrons in quantum dots interact via the long-range Coulomb interaction, and hence our model is a generalized or extended Hubbard model which includes both on-site and inter-site Coulomb interaction.  In addition, electrons in quantum dots could, in general, have distant neighbor hopping, not just nearest-neighbor hopping as in the minimal Hubbard model.  We therefore include both nearest-neighbor and next-nearest-neighbor hopping in the theory.  One other possible practical complication, which may be relevant to the experimental quantum dot arrays, is that each dot may have more than one relevant orbital level, making the system akin to an SU(2$n$) Hubbard model where $n$ is the number of orbitals (``quantum dot energy levels'') playing a role in each dot.\cite{OnufrievPRB1999}  In such a situation, the inter-site hopping process could involve inter-orbital hopping also.  We neglect this complication and consider a purely SU(2) system with each dot having just two spin states, assuming the higher orbital levels in each dot to be reasonably high in energy.  This is not an essential approximation, and is done to enable us to carry out our work completely analytically.  In any case, the neglect of higher orbital levels is a well-defined and well-controlled theoretical approximation since this can always be achieved experimentally by making each dot confinement potential sufficiently deep (and keeping the temperature sufficiently low) so that only the lowest orbital state in each dot is operational in the physics of the system.  The finite size Hubbard model we consider is therefore a generalization of the minimal Hubbard model, and includes both distant neighbor hopping and inter-site Coulomb interaction, but no higher orbital physics.

We also should mention here that although the quantum ferromagnetism discussed in our work is adiabatically connected to the Nagaoka ferromagnetism in the half-filled infinite-$U$ Hubbard model with one hole, there are important differences to keep in mind in order to avoid confusion and misunderstanding.  First, our system is a finite 2D plaquette (Fig. \ref{fig:geometries}) with 4 dots and 3-5 electrons whereas Nagaoka ferromagnetism is obviously a thermodynamic result.  Second, in our system the interaction could be large, but never infinite, since the infinite-$U$ limit is unphysical for actual quantum dots.  Third, our model being semi-realistic includes distant neighbor hopping and interaction, so we are considering a generalized and extended Hubbard model.  Fourth, our inter-site tunneling (i.e. the hopping parameter $t$) matrix element is negative, not positive as in the original work of Nagaoka.  Fifth, because of the small size of our system, one missing electron (i.e. a hole) corresponds to a finite hole density in contrast to the Nagaoka situation where the hole density is by definition zero (e.g. 3 electrons in a 2D square with 4 dots at the corners correspond to one hole in the system, but the hole density is 25\%!). Thus, the ferromagnetism we consider should perhaps be better called ``Nagaoka-type ferromagnetism'' rather than just Nagaoka ferromagnetism.  The really important point is, however, the fact that the quantum ferromagnetism we predict can be observed experimentally in already existing semiconductor quantum dot arrays.

The rest of this manuscript is organized as follows.  In sec. II, we investigate Nagaoka-type ferromagnetism by finding the ground states of three electrons in 4-dot plaquettes of various geometries. In sec. III, we repeat the calculations for a half-filled band (4 electrons) for the same geometries. In sec. IV, we look at the case of one hole in a 5-dot ring, and we summarize our results in sec. V.

\section{Three Electrons in Four Dots}

\subsection{General Model And Method}

\subsubsection{Hamiltonian}

We consider a single-band Hubbard model with onsite interaction energy $U_0$, long-range Coulomb interaction terms $V_{ij}$ and hopping terms $t_{ij}$. Thus the Hamiltonian is given by:
\begin{equation}
H=\sum_{i\ne j,\alpha}t_{ij}\,c_{i,\alpha}^\dagger c_{j,\alpha}+\sum_{i}U_0\,n_{i\uparrow}n_{i\downarrow}+\sum_{i\ne j}\frac{V_{ij}}{2}n_{i}n_{j}
\label{eqn:hgen}
\end{equation}

\begin{figure}[!htb]
\includegraphics[width=\columnwidth]{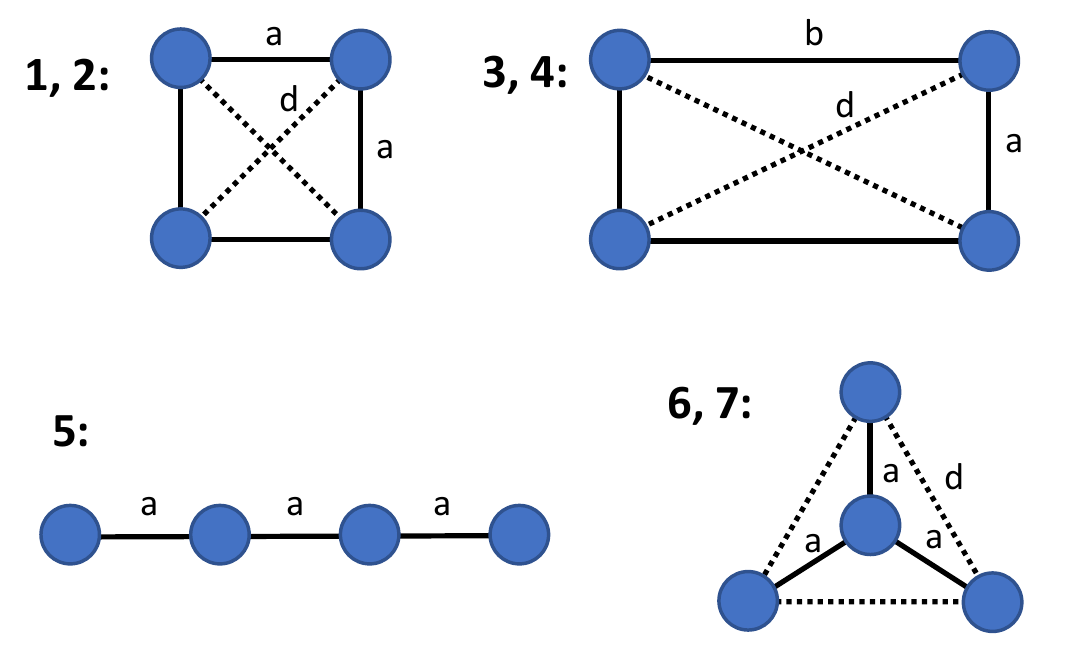}
\caption{A depiction of different 4-dot geometries, numbered as they appear in this work. Solid lines depict nearest-neighbor hopping terms, and dashed lines next nearest neighbor hopping terms, which we consider in some cases. In all cases long-range Coulomb interactions are included.}
\label{fig:geometries}
\end{figure}

Nagaoka's theorem predicts ferromagnetism in systems with one hole in a half-filled band with certain geometries where Nagaoka's condition holds. The simplest of these systems are a triangle or square plaquette of three or four sites. However, of particular importance is the sign of the product of hopping elements around loops $t_{12}t_{23}t_{31}$. In order for the Nagaoka condition to hold, quantities of this form must be positive; however, in reality, this sign is determined by the number of sites in the loop, and is negative for an odd number of sites. Thus a triangular plaquette with two electrons does not satisfy the Nagaoka condition, as must be the case since it is well known that the ground state of two electrons in any potential must necessarily be a singlet. Thus the addition of next nearest neighbor hopping terms (the dashed lines in fig. \ref{fig:geometries}) break the Nagaoka condition and can potentially destroy ferromagnetism if strong enough. It is interesting to derive a condition on the relative strengths of the hopping terms that determines whether ferromagnetism exists.

We consider four different geometries with 4 quantum dots: a square, a rectangle, a linear array, and Y-shaped plaquette, all with and without diagonal hopping terms where applicable. We note that only the first two satisfy the Nagaoka condition, and only in the absence of the diagonal hopping, as discussed above. We define $a$ to be the distance between nearest neighbors, along with $b>a$ in the case of the rectangle, and we define $d$ to be the distance between next nearest neighbors in each respective geometry. We define $V_r$ to be the Coulomb interaction energy between electrons separated by a distance $r$, and $t_r$ be the magnitude of the hopping strength between dots separated by a distance $r$. $U_0$ will be the onsite interaction energy as defined above. The bare parameters $V_r$ and $U_0$ are not important by themselves, but rather their differences are what affect the dynamics of the system, as a uniform shift in all these values will simply cause a constant shift in total energy, since the number of particles is conserved. Thus we will define new parameters $U$ and $V$ corresponding to the relevant energy differences, which vary for each geometry. We will also shift the total energy of the Hamiltonian by a constant such that the lowest energy configuration of electrons in the absence of tunneling is 0.


\subsubsection{Spin 3/2 States}

A system of three electrons can have either spin 1/2 or 3/2. To investigate the spin 3/2 states, we merely consider the case where all electrons are spin up, as all other states in the spin 3/2 quartet will be identical, aside from the value of $S_z$. We define the notation $\ket{d_1d_2d_3d_4}$ to be the state where the electron filling of dot $i$ is given by $d_i$, where $d_i\in\{0,\uparrow,\downarrow,\ud\}$. Since the Pauli exclusion principle forbids two spin up electrons from occupying the same orbital state, there are four possible spin 3/2 states for each value of $S_z$. For $S_z=3/2$, these are:
\begin{equation}
\begin{matrix}
\ket{\uparrow\,\uparrow\,\uparrow 0},&\ket{\uparrow\,\uparrow 0\uparrow},&\ket{\uparrow 0\uparrow\,\uparrow},&\ket{0\uparrow\,\uparrow\,\uparrow}
\end{matrix}
\label{eqn:basis3/2}
\end{equation}

The Hamiltonian is then constructed in this basis and diagonalized to find the eigenstates and energies. The lowest energy spin 3/2 state is compared to the lowest energy spin 1/2 state to detrmine whether the ground state is ferromagnetic.  Additionally, for comparison, we calculate the spin gap $\Delta$, defined to be the energy difference between the two lowest energy spin 3/2 states.

\subsubsection{Spin 1/2 States}

For the spin 1/2 state, we consider the case where two electrons are spin up and one is spin down. For configurations with at most one electron per site, this gives three states, one of which is part of the spin 3/2 quartet, and the other two of which have spin 1/2, as follows:
\begin{align}
\ket{\psi_{3/2}}&=\frac{1}{\sqrt{3}}\big(\ket{\uparrow\uparrow\downarrow}+\ket{\uparrow\downarrow\uparrow}+\ket{\downarrow\uparrow\uparrow}\big)\nonumber\\
\ket{\psi_{1/2}^+}&=\frac{1}{\sqrt{3}}\big(e^{\frac{2\pi i}{3}}\ket{\uparrow\uparrow\downarrow}+\ket{\uparrow\downarrow\uparrow}+e^{\frac{-2\pi i}{3}}\ket{\downarrow\uparrow\uparrow}\big)\nonumber\\
\ket{\psi_{1/2}^-}&=\frac{1}{\sqrt{3}}\big(e^{\frac{-2\pi i}{3}}\ket{\uparrow\uparrow\downarrow}+\ket{\uparrow\downarrow\uparrow}+e^{\frac{2\pi i}{3}}\ket{\downarrow\uparrow\uparrow}\big)
\label{eqn:sz1/2}
\end{align}

Define a matrix $M$ such that
\begin{equation}
\begin{pmatrix}\ket{\psi_{1/2}^+}\\\ket{\psi_{1/2}^-}\end{pmatrix}=M\begin{pmatrix}\ket{\uparrow\uparrow\downarrow}\\\ket{\uparrow\downarrow\uparrow}\\\ket{\downarrow\uparrow\uparrow}\end{pmatrix}
\end{equation}

which can be obtained simply by reading off the coefficients of eq. (\ref{eqn:sz1/2}). Then we have a total of 8 low-energy spin 1/2 states with $S_z=1/2$:
\begin{equation}
\begin{matrix}
\ket{\psi^+_1\psi^+_2\psi^+_30},&\ket{\psi^+_1\psi^+_20\psi^+_3},&\ket{\psi^+_10\psi^+_2\psi^+_3},&\ket{0\psi^+_1\psi^+_2\psi^+_3},\\&&&\\
\ket{\psi^-_1\psi^-_2\psi^-_30},&\ket{\psi^-_1\psi^-_20\psi^-_3},&\ket{\psi^-_10\psi^-_2\psi^-_3},&\ket{0\psi^-_1\psi^-_2\psi^-_3}
\end{matrix}
\label{eqn:s12basis3e4d}
\end{equation}

Here $\psi^i_j$ refers to the state of the $j$th spin of $\ket{\psi^i_{1/2}}$ defined as in eq. (\ref{eqn:sz1/2}). For example, the state $\ket{\psi^+_10\psi^+_2\psi^+_3}=\frac{1}{\sqrt{3}}(e^{\frac{2\pi i}{3}}c_{1\uparrow}^\dagger c_{3\uparrow}^\dagger c_{4\downarrow}^\dagger+c_{1\uparrow}^\dagger c_{3\downarrow}^\dagger c_{4\uparrow}^\dagger+e^{\frac{-2\pi i}{3}}c_{1\downarrow}^\dagger c_{3\uparrow}^\dagger c_{4\uparrow}^\dagger)\ket{0}$. There are also 12 high energy states, corresponding to all permutations of $\ket{\ud\uparrow0\,0}$. These states only affect the energies to order $t^2/U$. Since Nagaoka's theorem applies only in the infinite $U$ limit, we will initially consider only the low energy states, and afterward calculate corrections to order $t^2/U$.

Nagaoka ferromagnetism occurs because as a hole tunnels around a loop, it causes the other electron spins in the loop to be cyclically shifted one position. In the ferromagnetic state, all spins point in the same direction, and thus cycling them does not change the spin configuration. At a lower total spin, however, there is a mixture of up and down spins, and thus cycling them will have some effect such as rotating one spin configuration into another or adding a phase, which can potentially increase the energy of the state with lower total spin. In our calculation, we see this effect when calculating the matrix elements of $H$ between states where one electron has tunneled. If the two dots where the tunneling occurred are in consecutive order, then the spins remain in the same order, and the matrix element is given by the corresponding term in the Hamiltonian, as in the following example:
\begin{equation}
\bra{s_1s_2s_30}H\ket{s_1's_2'0s_3'}=-t\delta_{s_1s_1'}\delta_{s_2s_2'}\delta_{s_3s_3'}
\end{equation}

and thus matrix elements between $\psi^i$ can be found via:
\begin{equation}
\bra{\psi^i_1\psi^i_2\psi^i_30}H\ket{\psi^j_1\psi^j_20\psi^j_3}=\Big(M^*(-t)M^T\Big)_{ij}=-t\delta_{ij}
\end{equation}

and similarly for all other states of this form. However, if the dots are not in consecutive order, such as for example hopping between dots 1 and 4, then the spins can potentially be rearranged:
\begin{equation}
\bra{s_1s_2s_30}H\ket{0s_1's_2's_3'}=-t\delta_{s_2s_1'}\delta_{s_3s_2'}\delta_{s_1s_3'}
\end{equation}

and therefore:
\begin{flalign}
&\bra{\psi^i_1\psi^i_2\psi^i_30}H\ket{0\psi^j_1\psi^j_2\psi^j_3}=-t\Bigg[M^*\begin{pmatrix}0&1&0\\0&0&1\\1&0&0\end{pmatrix}M^T\Bigg]_{ij}&&\nonumber\\
&=\begin{pmatrix}-te^{\frac{-2\pi i}{3}}&0\\0&-te^{\frac{2\pi i}{3}}\end{pmatrix}_{ij}&&
\label{eqn:cyclemix}
\end{flalign}

\subsubsection{Finite $U$ Corrections}

For several of the geometries, we also determine the leading order corrections to $E_{1/2}$ for $U\gg t$ but not infinite. This is done using perturbation theory, but is complicated by the fact that the spin 0 states are often degenerate. We determine the matrix elements of $H$ between the lowest energy spin 0 states, which we denote $\ket{\Psi_{1/2}^i}$ and the high energy $(2,1,0,0)$ states, which we denote $\ket{\Phi^i}$ and order as follows:

\begin{equation}
\begin{matrix}
\ket{\ud 0\uparrow 0},&\ket{0\ud 0\uparrow},&\ket{\uparrow 0\ud 0},&\ket{0\uparrow 0\ud},\\\ket{\ud\uparrow 0\,0},&\ket{\ud 0\,0\uparrow},&\ket{0\ud\uparrow 0},&\ket{\uparrow\ud 0\,0},\\\ket{0\,0\ud\uparrow},&\ket{0\uparrow\ud 0},&\ket{\uparrow 0\,0\ud},&\ket{0\,0\uparrow\ud},
\end{matrix}
\end{equation}


We define the matrices $T$ and $\Lambda$ as follows:
\begin{align}
T_{ij}&=\bra{\Phi^i}H\ket{\Psi_{1/2}^j}\\
\Lambda_{ij}&=\bra{\Phi^i}H\ket{\Phi^j}
\end{align}

Note that $\Lambda$ is diagonal to leading order in $t/U$, and is given simply by the energies of $\ket{\Phi^i}$. Then the corrections to the singlet state energies to order $t^2/U$ are given by the eigenvalues of the matrix $-T^\dagger\Lambda^{-1}T$.

\subsection{Ground State Calculations}

\subsubsection{Square with no Diagonal Hopping}

We initially consider a system of four dots in a square, where $t_{ij}$ and $V_{ij}$ are given as follows:
\begin{equation}
t_{ij}=\begin{cases}
-t_a&\text{if }i-j=\pm1\mod 4\\
0&\text{otherwise}
\end{cases}
\end{equation}
\begin{equation}
V_{ij}=\begin{cases}
V_a&\text{if }i-j=\pm1\mod 4\\
V_d&\text{if }i-j=2\mod 4
\end{cases}
\end{equation}

Up to symmetry, three different electron configurations are possible:
\begin{align}
&(1,1,1,0)\;\text{ with energy: }\;2V_a+V_d\nonumber\\
&(2,0,1,0)\;\text{ with energy: }\;U_0+2V_d\nonumber\\
&(2,1,0,0)\;\text{ with energy: }\;U_0+2V_a
\label{eqn:sqconfig}
\end{align}

We shift the total energy of the Hamiltonian by a constant amount $2V_a+V_d$, and define $U$ and $V$ as:
\begin{align}
&U\equiv U_0-2V_a+V_d\nonumber\\
&V\equiv V_a-V_d
\label{eqn:uvdefsq}
\end{align}

so that the energies of the three electron configurations in eq. (\ref{eqn:sqconfig}) become $0$, $U$, and $U+2V$ respectively. Then the spin 3/2 Hamiltonian in the basis given by eq. (\ref{eqn:basis3/2}) is:
\begin{equation}
H_{3/2}=-t_a\begin{pmatrix}
0&1&0&1\\
1&0&1&0\\
0&1&0&1\\
1&0&1&0
\end{pmatrix}
\end{equation}

which has ground state $\Psi_{3/2}=\frac{1}{2}(1\,1\,1\,1)^T$ and energy $E_{3/2}=-2t_a$. The first excited spin 3/2 state has energy 0, so the spin gap is $\Delta=2t_a$.

We now find the spin 1/2 Hamiltonian. From eq. (\ref{eqn:cyclemix}), a phase is introduced when tunneling the hole around the loop. Thus the spin 1/2 Hamiltonian is given by a block diagonal matrix consisting of two blocks, corresponding to $\psi_{1/2}^\pm$ as defined in eq. (\ref{eqn:sz1/2}):
\begin{equation}
H_{1/2}^{\pm}=-t_a\begin{pmatrix}
0&1&0&e^{\mp\frac{2\pi i}{3}}\\
1&0&1&0\\
0&1&0&1\\
e^{\pm\frac{2\pi i}{3}}&0&1&0
\end{pmatrix}
\end{equation}

which has ground states given by:
\begin{align}
\Psi_{1/2}^\pm=\frac{1}{2}&\bigg[\ket{\psi^\pm_1\psi^\pm_2\psi^\pm_30}+e^{\pm\frac{\pi i}{6}}\ket{\psi^\pm_1\psi^\pm_20\psi^\pm_3}\nonumber\\&+e^{\pm\frac{\pi i}{3}}\ket{\psi^\pm_10\psi^\pm_2\psi^\pm_3}\pm i\ket{0\psi^\pm_1\psi^\pm_2\psi^\pm_3}\bigg]
\end{align}

with energy $E^\pm_{1/2}=-\sqrt{3}t_a$. Thus in the infinite $U$ limit, the system exhibits ferromagnetism, since the spin 3/2 state has lower energy.

\begin{figure}[!htb]
	\includegraphics[width=\columnwidth]{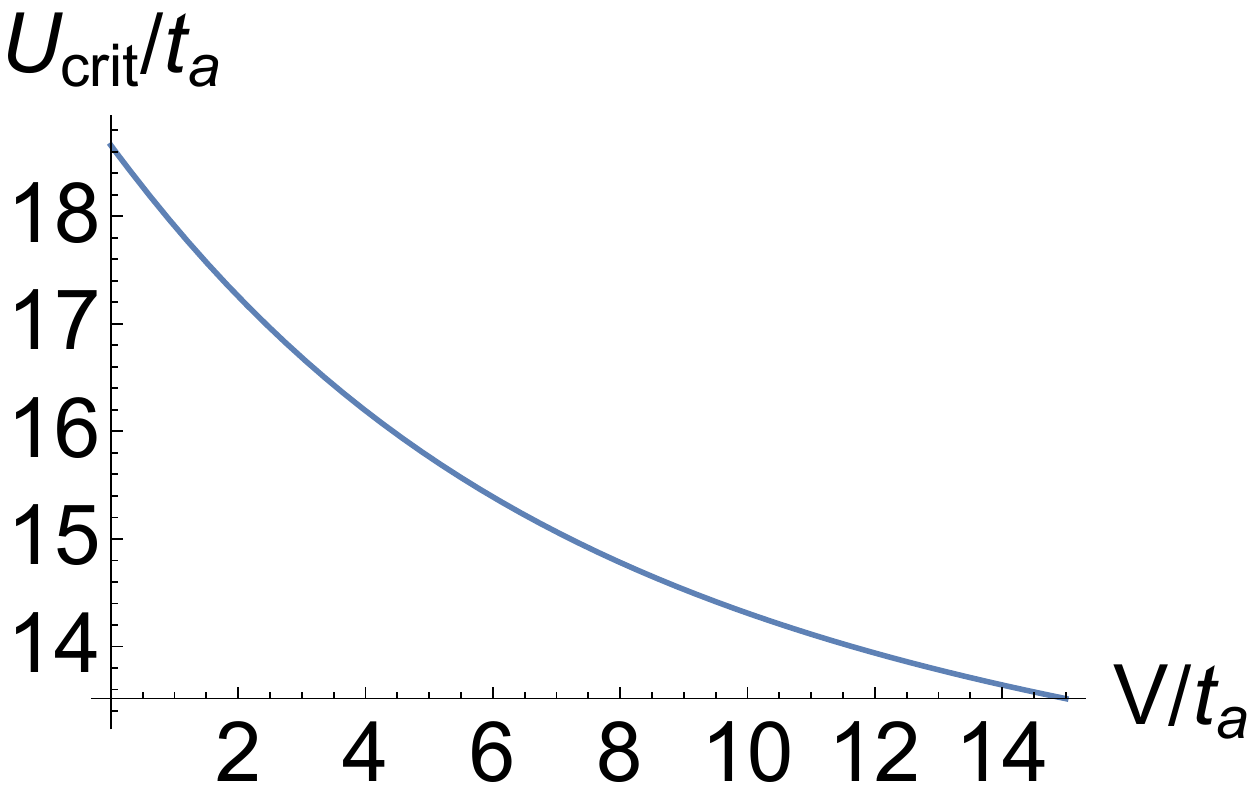}
	\caption{$U_{\text{crit}}$ versus $V$ for three electrons in a four-dot square configuration. Here $U$ and $2V$ are defined as in eq. (\ref{eqn:uvdefsq}).}
\end{figure}

\begin{figure}[!htb]
	\includegraphics[width=\columnwidth]{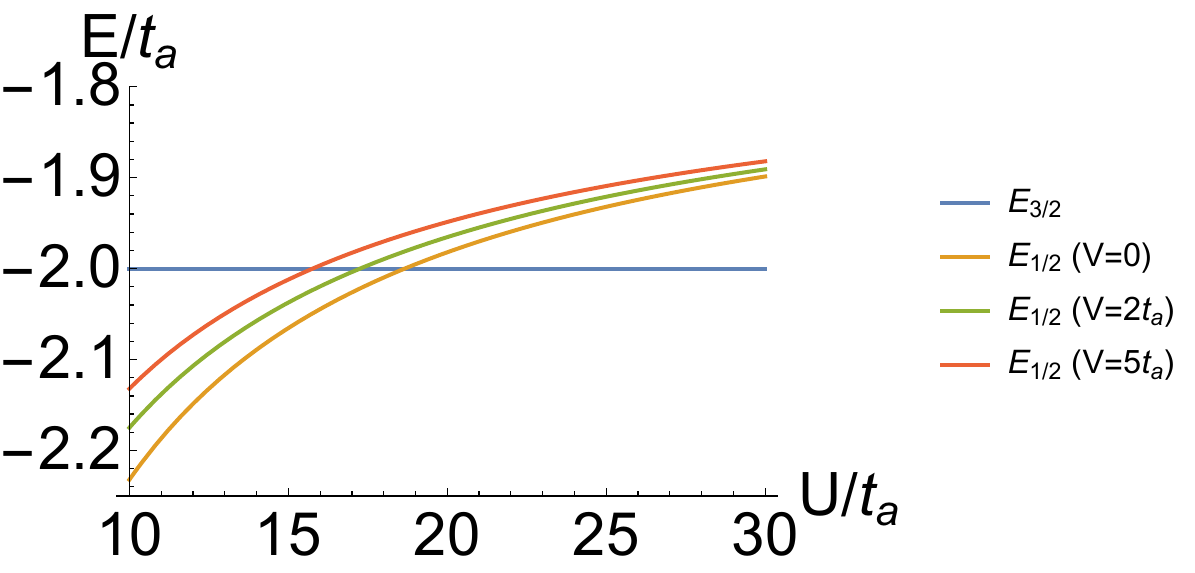}
	\caption{$E_{3/2}$ and $E_{1/2}$ versus $U$ for different values of $V$ for three electrons in a four-dot square configuration. The point where $E_{3/2}$ and $E_{1/2}$ cross is $U_{\text{crit}}$.}
\end{figure}

We also determine the finite $U$ corrections to $E_{1/2}$. Since there are two degenerate spin 1/2 states, $-T^\dagger\Lambda^{-1}T$ is a $2\times 2$ matrix, given by:
\begin{equation}
-T^\dagger\Lambda^{-1}T=\Big[-3\frac{t_a^2}{U}-2\frac{t_a^2}{U+2V}\Big]\begin{pmatrix}1&0\\0&1\end{pmatrix}
\end{equation}

Hence we find that the $\Psi_{1/2}^\pm$ degeneracy remains unbroken, and the spin 1/2 ground state energy is given by:
\begin{equation}
E_{1/2}=-\sqrt{3}t_a-3\frac{t_a^2}{U}-2\frac{t_a^2}{U+2V}+O\Big(\frac{t_a^3}{U^2}\Big)
\label{eqn:e12sq}
\end{equation}

Then for $V\rightarrow0$, we recover a correction of $-5t_a^2/U$, agreeing with the result given in Ref. \onlinecite{DehollainARXIV2019}. Using this result, we can derive the value $U_{\text{crit}}$ (to first order in $t_a/U$) which marks the transition between the ferromagnetic and antiferromagnetic phases:
\begin{align}
U_{\text{crit}}&=\frac{1}{2(2-\sqrt{3})}\Bigg[-2(2-\sqrt{3})V+5t_a+\nonumber\\&\sqrt{(2(2-\sqrt{3})V-5t_a)^2+24(2-\sqrt{3})Vt_a}\,\Bigg]
\end{align}

For $V\rightarrow 0$, this gives $U_{\text{crit}}=5t_a/(2-\sqrt{3})\approx18.7t_a$.

\subsubsection{Square with Diagonal Hopping}

We now investigate how diagonal hopping terms effect the system. We use the same square configuration of four dots, but now add extra hopping terms $t_{13}=t_{31}=t_{42}=t_{24}=-t_d$. We again define $U$ and $V$ as in equation (\ref{eqn:uvdefsq}). The analysis for the spin 3/2 states is similar to above, except there are now extra matrix elements corresponding to $t_d$. These will be positive rather than negative as an extra minus sign is introduced due to Fermi statistics, since diagonal tunneling essentially exchanges two electrons. Then the spin 3/2 Hamiltonian is given as follows:
\begin{equation}
H_{3/2}=\begin{pmatrix}
0&-t_a&t_d&-t_a\\
-t_a&0&-t_a&t_d\\
t_d&-t_a&0&-t_a\\
-t_a&t_d&-t_a&0
\end{pmatrix}
\end{equation}

which has ground state $\Psi_{3/2}=\frac{1}{2}(1\,1\,1\,1)^T$ and energy $E_{3/2}=-2t_a+t_d$. The first excited state has energy $-t_d$, so the spin gap is $\Delta=2t_a-2t_d$.

\begin{figure}[!htb]
	\includegraphics[width=\columnwidth]{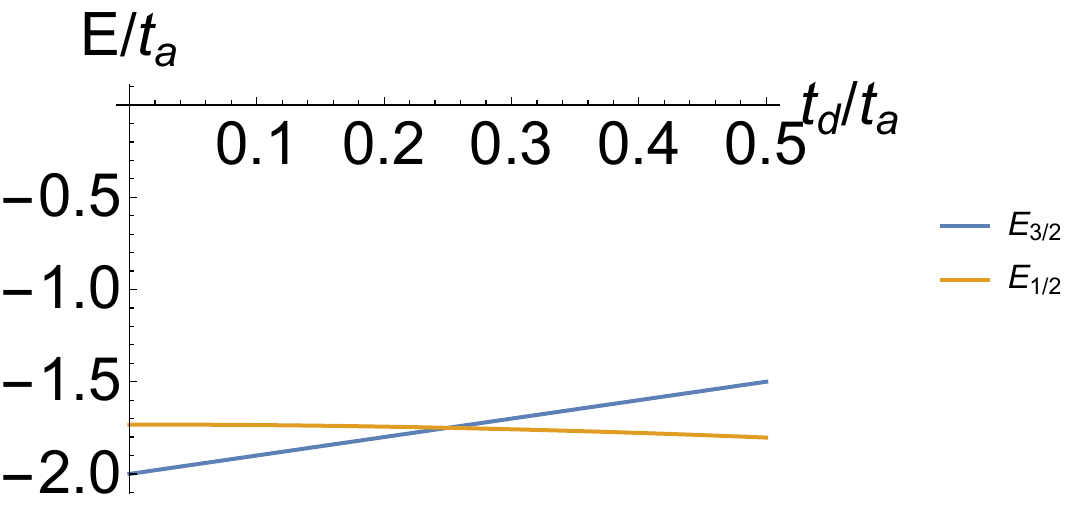}
	\caption{Plot of $E_{3/2}$ and $E_{1/2}$ versus $t_d/t_a$ for three electrons in a four-dot square configuration with diagonal hopping in the infinite $U$ limit. We see that ferromagnetism is only possible for $t_d<t_a/4$.}
\end{figure}

The analysis for the spin 1/2 states is also similar to the square model, with again the only difference in the infinite $U$ limit being the diagonal hopping terms $t_d$. Then a calculation similar to eq. (\ref{eqn:cyclemix}) yields:

\begin{align}
\bra{\psi^i_1\psi^i_2\psi^i_30}H\ket{\psi^j_10\psi^j_2\psi^j_3}&=\begin{pmatrix}0&t_de^{\frac{-2\pi i}{3}}\\t_de^{\frac{2\pi i}{3}}&0\end{pmatrix}_{ij}\nonumber\\
\bra{\psi^i_1\psi^i_20\psi^i_3}H\ket{0\psi^j_1\psi^j_2\psi^j_3}&=\begin{pmatrix}0&t_de^{\frac{2\pi i}{3}}\\t_de^{\frac{-2\pi i}{3}}&0\end{pmatrix}_{ij}
\end{align}

Thus diagonal hopping rotates $\ket{\psi_{1/2}^+}$ into $\ket{\psi_{1/2}^-}$ and vice versa. Then $H_{1/2}$ is no longer block-diagonal, and is given by:
\begin{widetext}
	\begin{equation}
	H_{1/2}=\begin{pmatrix}
	0&-t_a&0&-t_ae^{\frac{-2\pi i}{3}}&0&0&t_de^{\frac{-2\pi i}{3}}&0\\
	-t_a&0&-t_a&0&0&0&0&t_de^{\frac{2\pi i}{3}}\\
	0&-t_a&0&-t_a&t_de^{\frac{-2\pi i}{3}}&0&0&0\\
	-t_ae^{\frac{2\pi i}{3}}&0&-t_a&0&0&t_de^{\frac{2\pi i}{3}}&0&0\\
	0&0&t_de^{\frac{2\pi i}{3}}&0&0&-t_a&0&-t_ae^{\frac{2\pi i}{3}}\\
	0&0&0&t_de^{\frac{-2\pi i}{3}}&-t_a&0&-t_a&0\\
	t_de^{\frac{2\pi i}{3}}&0&0&0&0&-t_a&0&-t_a\\
	0&t_de^{\frac{-2\pi i}{3}}&0&0&-t_ae^{\frac{-2\pi i}{3}}&0&-t_a&0
	\end{pmatrix}
	\end{equation}
\end{widetext}

which has two degenerate ground states with energy $E_{1/2}=-\sqrt{3t_a^2+t_d^2}$. Thus, in the infinite $U$ limit, $E_{3/2}<E_{1/2}$ as long as $t_d<t_a/4$, and thus ferromagnetism only exists for $t_d<t_a/4$.

\subsubsection{Rectangle with no Diagonal Hopping}

We now model a rectangular configuration of four dots. This will be similar to the square model, except $t_{ij}$ and $V_{ij}$ are given by:
\begin{equation}
t_{ij}=\begin{cases}
-t_a&\text{if }\{i,j\}=\{1,2\}\text{ or }\{3,4\}\\
-t_b&\text{if }\{i,j\}=\{2,3\}\text{ or }\{1,4\}\\
0&\text{otherwise}\\
\end{cases}
\label{eqn:trec}
\end{equation}
\begin{equation}
V_{ij}=\begin{cases}
V_a&\text{if }\{i,j\}=\{1,2\}\text{ or }\{3,4\}\\
V_b&\text{if }\{i,j\}=\{2,3\}\text{ or }\{1,4\}\\
V_d&\text{if }i-j=\pm 2\\
\end{cases}
\end{equation}

Without loss of generality, we will assume $b>a$, and thus $t_a>t_b$ and $V_a>V_b$. We note that up to symmetry the following four electron configurations are possible:
\begin{align}
&(1,1,1,0)\;\text{ with energy: }\;V_a+V_b+V_d\nonumber\\
&(2,0,1,0)\;\text{ with energy: }\;U_0+2V_d\nonumber\\
&(2,0,0,1)\;\text{ with energy: }\;U_0+2V_b\nonumber\\
&(2,1,0,0)\;\text{ with energy: }\;U_0+2V_a
\label{eqn:recconfig}
\end{align}

We shift the total energy by $V_a+V_b+V_d$, and define $U$, $V$, and $W$ as:
\begin{align}
U&\equiv U_0-V_a-V_b+V_d\nonumber\\
V&\equiv V_a-V_d\nonumber\\
W&\equiv V_b-V_d
\label{eqn:uvdefrec}
\end{align}

so that the energies of the electron configurations in eq. (\ref{eqn:recconfig}) become $0$, $U$, $U+2W$, $U+2V$ respectively. The analysis for the spin 3/2 states is identical to the square model, except that care must be taken to distinguish between $t_a$ and $t_b$. Thus we construct the Hamiltonian:
\begin{equation}
H_{3/2}=\begin{pmatrix}
0&-t_a&0&-t_b\\
-t_a&0&-t_b&0\\
0&-t_b&0&-t_a\\
-t_b&0&-t_a&0
\end{pmatrix}
\end{equation}

which has ground state $\Psi_{3/2}=\frac{1}{2}(1\,1\,1\,1)^T$ and energy $E_{3/2}=-t_a-t_b$. The first excited state has energy $-t_a+t_b$, so the spin gap is $\Delta=2t_b$.

\begin{figure}[!htb]
	\includegraphics[width=\columnwidth]{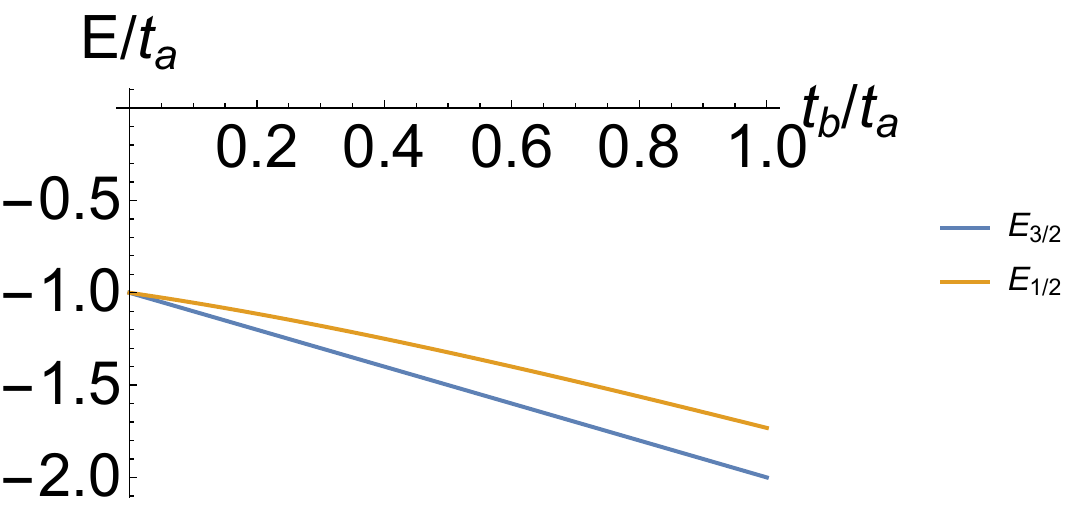}
	\caption{Plot of $E_{3/2}$ and $E_{1/2}$ versus $t_b/t_a$ for three electrons in a four-dot rectangular configuration with no diagonal hopping in the infinite $U$ limit.}
\end{figure}

The analysis for the spin 1/2 states is also similar to the square model, with again the only difference in the infinite $U$ limit being the the second hopping strength $t_b$. Then the spin 1/2 Hamiltonian is given by:
\begin{equation}
H_{1/2}^{\pm}=\begin{pmatrix}
0&-t_a&0&-t_be^{\mp\frac{2\pi i}{3}}\\
-t_a&0&-t_b&0\\
0&-t_b&0&-t_a\\
-t_be^{\pm\frac{2\pi i}{3}}&0&-t_a&0
\end{pmatrix}
\end{equation}

which has energy $E_{1/2}^\pm=-\sqrt{t_a^2+t_at_b+t_b^2}$, and ground state given by:
\begin{align}
\Psi_{1/2}^\pm=\frac{1}{2}\bigg[&\ket{\psi^\pm_1\psi^\pm_2\psi^\pm_30}+e^{\pm i\varphi}\ket{\psi^\pm_1\psi^\pm_20\psi^\pm_3}\nonumber\\&+e^{\pm i\frac{\pi}{3}}\ket{\psi^\pm_10\psi^\pm_2\psi^\pm_3}+e^{\pm i(\varphi+\frac{\pi}{3})}\ket{0\psi^\pm_1\psi^\pm_2\psi^\pm_3}\bigg]
\end{align}

where $\varphi\equiv\arctan\frac{\sqrt{3}t_b}{2t_a+t_b}$. Thus, three electrons in four dots arranged in a rectangular configuration will exhibit ferromagnetism for large $U$, regardless of the ratio of $t_a$ and $t_b$. This is assuming that there is no diagonal hopping, an assumption that may break down for extreme ratios of $t_a$ to $t_b$. 

\begin{figure}[!htb]
	\includegraphics[width=\columnwidth]{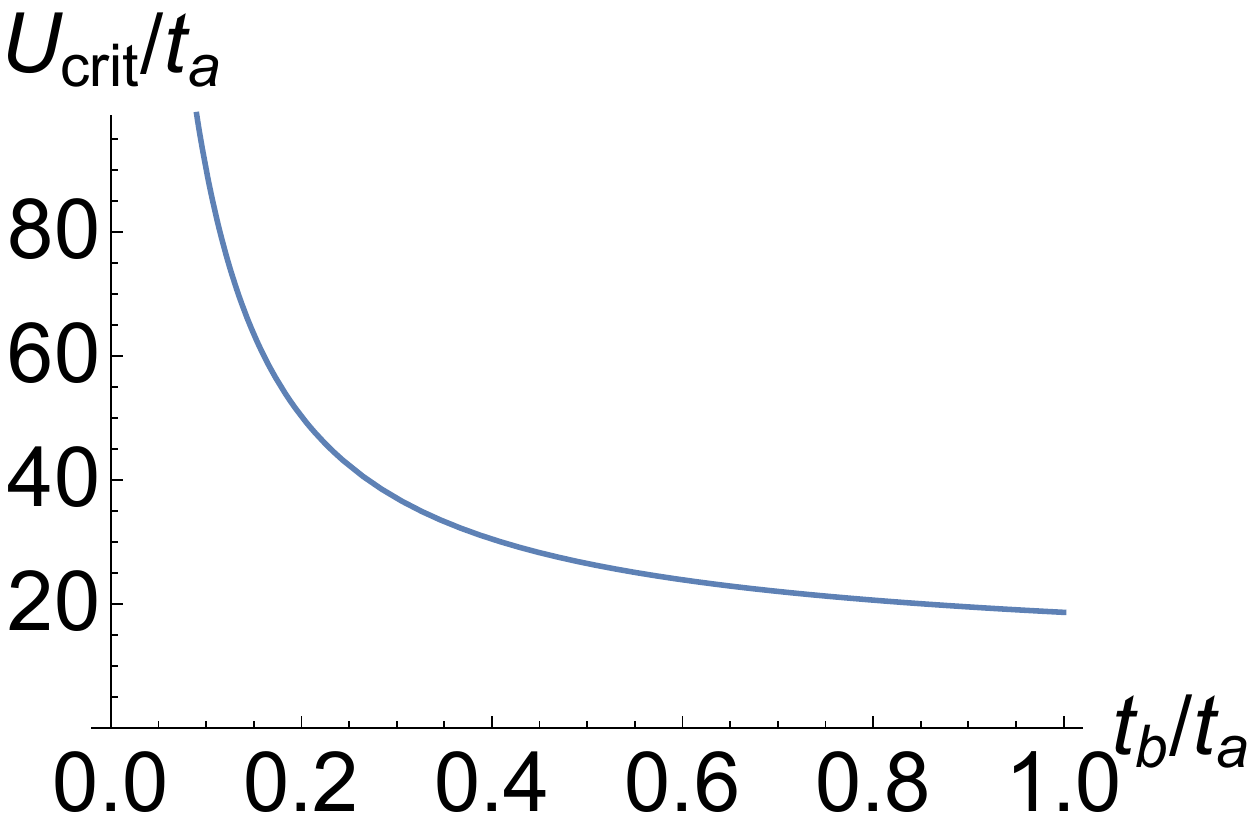}
	\caption{Plot of $U_{\text{crit}}$ versus $t_b/t_a$ for three electrons in a four-dot rectangular configuration with $V=W=0$.}
\end{figure}

\begin{figure}[!htb]
	\includegraphics[width=\columnwidth]{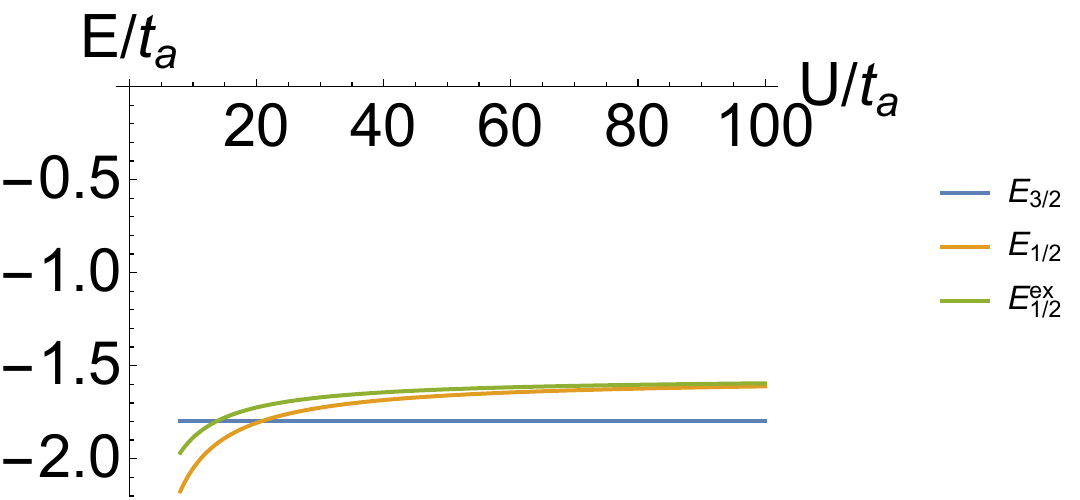}		\includegraphics[width=\columnwidth]{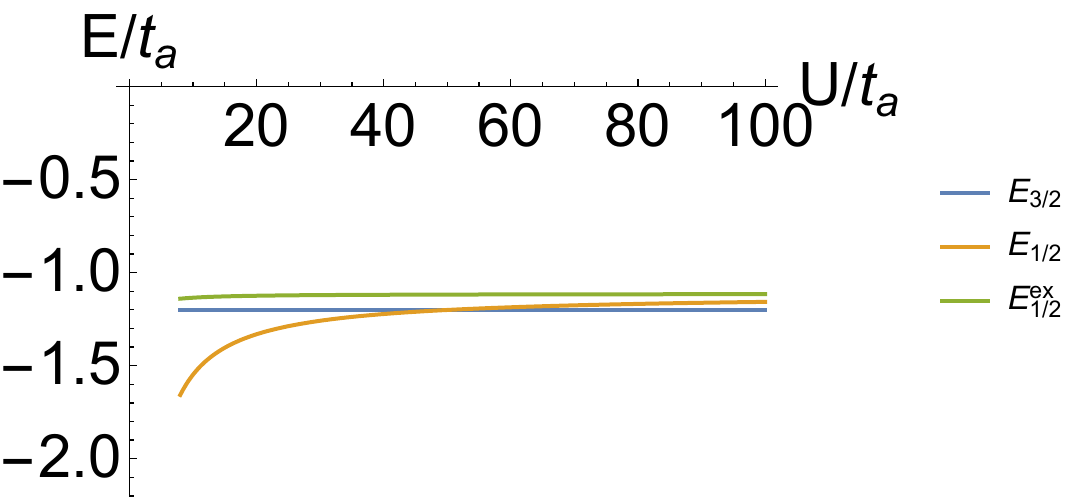}
	\caption{Plot of $E_{3/2}$, $E_{1/2}$, and the nearly-degenerate excited state energy $E_{1/2}^{\text{ex}}$ versus $U$ for three electrons in a four-dot rectangular configuration with no diagonal hopping with $t_b/t_a=.8$ (Top) and $t_b/t_a=.2$ (Bottom). Here $V=W=0$.}
\end{figure}

The procedure for calculating the finite $U$ corrections to $E_{1/2}^\pm$ is also similar to the square model. We calculate $-T^\dagger\Lambda^{-1}T$ like before, obtaining:
\begin{align}
&-T^\dagger\Lambda^{-1}T\nonumber\\
=&\,\frac{-t_a^2-t_at_b-t_b^2}{U}\begin{pmatrix}1&e^{\frac{-i\pi}{3}-i\varphi}\cos 3\varphi\\e^{\frac{i\pi}{3}+i\varphi}\cos 3\varphi&1\end{pmatrix}\nonumber\\
&-\,\frac{t_a^2}{U+2W}\begin{pmatrix}1&e^{\frac{-i\pi}{3}-i\varphi}\cos \varphi\\e^{\frac{i\pi}{3}+i\varphi}\cos \varphi&1\end{pmatrix}\nonumber\\
&-\,\frac{t_b^2}{U+2V}\begin{pmatrix}1&\!\!\!\!\!\!\!\!-e^{\frac{-i\pi}{3}-i\varphi}\cos (\varphi-\frac{\pi}{3})\\-e^{\frac{i\pi}{3}+i\varphi}\cos (\varphi-\frac{\pi}{3})&\!\!\!\!\!\!\!\!1\end{pmatrix}\nonumber\\
\end{align}

The off-diagonal terms break the $\ket{\Psi_{1/2}^+},\ket{\Psi_{1/2}^-}$ degeneracy, with the lower energy state given by:
\begin{equation}
\ket{\Psi_{1/2}}=\frac{1}{\sqrt{2}}\Big[\ket{\Psi_{1/2}^+}+e^{\frac{i\pi}{3}+i\varphi}\ket{\Psi_{1/2}^-}\Big]
\end{equation}

and thus, the energy of the lowest energy state is:
\begin{align}
E_{1/2}=-\sqrt{t_a^2+t_at_b+t_b^2}-\,\frac{t_a^2+t_at_b+t_b^2}{U}(1+\cos 3\varphi)\nonumber\\-\,\frac{t_a^2}{U+2W}(1+\cos\varphi)-\,\frac{t_b^2}{U+2V}(1-\cos(\varphi-\frac{\pi}{3}))
\end{align}

\subsubsection{Rectangle with Diagonal Hopping}

We now address the case of diagonal hopping in a rectangular system. We define $t_a$ and $t_b$ as in eq. (\ref{eqn:trec}), and let the diagonal hopping term be given by $t_d$. We assume $t_a>t_b>t_d$. We shift the total energy by $V_a+V_b+V_d$, as in the rectangular case, and define $U$, $V$, $W$ as in equation (\ref{eqn:uvdefrec}).

The analysis for the spin 3/2 states is similar to above. Thus we construct the Hamiltonian:

\begin{equation}
H_{3/2}=\begin{pmatrix}
0&-t_a&t_d&-t_b\\
-t_a&0&-t_b&t_d\\
t_d&-t_b&0&-t_a\\
-t_b&t_d&-t_a&0
\end{pmatrix}
\end{equation}

which has ground state $\Psi_{3/2}=\frac{1}{2}(1\,1\,1\,1)^T$ and energy $E_{3/2}=-t_a-t_b+t_d$. The first excited state has energy $-t_a+t_b-t_d$, so the spin gap is $\Delta=2t_b-2t_d$.

The analysis for the spin 1/2 states is also similar to above. Then $H_{1/2}$ is given by:
\begin{widetext}
	\begin{equation}
	H_{1/2}=\begin{pmatrix}
	0&-t_a&0&-t_be^{\frac{-2\pi i}{3}}&0&0&t_de^{\frac{-2\pi i}{3}}&0\\
	-t_a&0&-t_b&0&0&0&0&t_de^{\frac{2\pi i}{3}}\\
	0&-t_b&0&-t_a&t_de^{\frac{-2\pi i}{3}}&0&0&0\\
	-t_be^{\frac{2\pi i}{3}}&0&-t_a&0&0&t_de^{\frac{2\pi i}{3}}&0&0\\
	0&0&t_de^{\frac{2\pi i}{3}}&0&0&-t_a&0&-t_be^{\frac{2\pi i}{3}}\\
	0&0&0&t_de^{\frac{-2\pi i}{3}}&-t_a&0&-t_b&0\\
	t_de^{\frac{2\pi i}{3}}&0&0&0&0&-t_b&0&-t_a\\
	0&t_de^{\frac{-2\pi i}{3}}&0&0&-t_be^{\frac{-2\pi i}{3}}&0&-t_a&0
	\end{pmatrix}
	\end{equation}
\end{widetext}

which has a nondegenerate ground state with energy $E_{1/2}=-\sqrt{t_a^2+t_b^2+t_d^2+t_at_b+t_at_d-t_bt_d}$. From this, it is easy to show that in the infinite $U$ limit, $E_{3/2}<E_{1/2}$ as long as $t_d<t_at_b/(3t_a+t_b)$.

\subsubsection{Linear Array of Four Dots}

We also model a linear array of four dots. This will be similar to the square model, except $t_{14}=t_{41}=0$, and $V_{ij}$ is given by:
\begin{equation}
V_{ij}=\begin{cases}
V_a&\text{if }i-j=\pm 1\\
V_{2a}&\text{if }i-j=\pm 2\\
V_{3a}&\text{if }i-j=\pm 3
\end{cases}
\end{equation}

We note that up to symmetry, the following electron configurations are possible:
\begin{align}
&(1,1,0,1)\;\text{ with energy: }\;V_a+V_{2a}+V_{3a}\nonumber\\
&(1,1,1,0)\;\text{ with energy: }\;2V_a+V_{2a}\nonumber\\
&(2,0,0,1)\;\text{ with energy: }\;U_0+2V_{3a}\nonumber\\
&(2,0,1,0)\;\text{ with energy: }\;U_0+2V_{2a}\nonumber\\
&(2,1,0,0)\;\text{ with energy: }\;U_0+2V_a
\label{eqn:lineconfig}
\end{align}

We shift the total energy by $V_a+V_{2a}+V_{3a}$, and define $U$, $V$, and $W$ as:
\begin{align}
U&\equiv U_0-V_a-V_{2a}+V_{3a}\nonumber\\
V&\equiv V_a-V_{3a}\nonumber\\
W&\equiv V_{2a}-V_{3a}
\end{align}

so that the energies of the electron configurations in eq. (\ref{eqn:lineconfig}) become $0$, $V$, $U$, $U+2W$, $U+2V$ respectively. The analysis for the spin 3/2 states is identical to the square model, except that some states have an extra energy $V$, and no hopping is permitted between dots 1 and 4. Thus we construct the Hamiltonian:
\begin{equation}
H_{3/2}=\begin{pmatrix}
V&-t_a&0&0\\
-t_a&0&-t_a&0\\
0&-t_a&0&-t_a\\
0&0&-t_a&V
\end{pmatrix}
\end{equation}

which has a nondegenerate ground state with energy:
\begin{equation}
E_{3/2}=(V-t_a-\sqrt{(V+t_a)^2+4t_a^2})/2
\end{equation}

and ground state given by:
\begin{equation}
\Psi_{3/2}=\frac{1}{\sqrt{2}\sqrt{1+\frac{(V-E_{3/2})^2}{t_a^2}}}\begin{pmatrix}1\\(V-E_{3/2})/t_a\\(V-E_{3/2})/t_a\\1\end{pmatrix}
\label{eqn:abdefref}
\end{equation}

For convenience, we define $A(V,t)$ and $B(V,t)$ from eq. (\ref{eqn:abdefref}) above such that $\Psi_{3/2}=(A\;B\;B\;A)^T$. The first excited state has energy $(V+t_a-\sqrt{(V-t_a)^2+4t_a^2})/2$, and so the spin gap is given by the difference of this energy and $E_{3/2}$.

In the square model without diagonal hopping, the only difference between the spin 3/2 and spin 1/2 subspaces in the infinite $U$ limit is in the hopping term between dots 1 and 4. Since this term no longer exists in the linear model, we find that the spin 1/2 Hamiltonian is simply two exact copies of the spin 3/2 Hamiltonian, $H_{1/2}^\pm=H_{3/2}$, and thus the ground state energy $E_{1/2}^\pm=E_{3/2}$, as well. Thus for finite $U$, the system cannot exhibit ferromagnetism, since the finite $U$ corrections will lower the energy of the spin 1/2 states.

\begin{figure}[!htb]
	\includegraphics[width=\columnwidth]{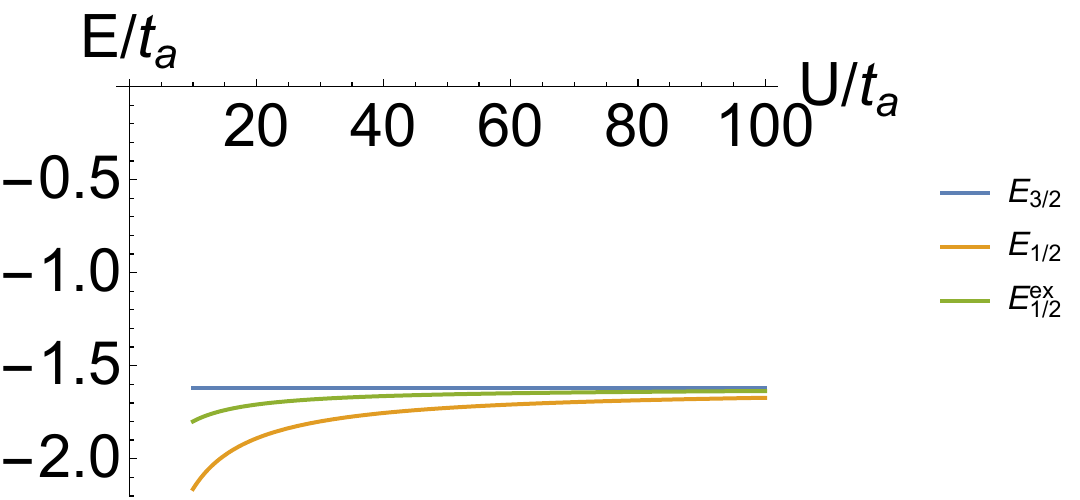}
	\includegraphics[width=\columnwidth]{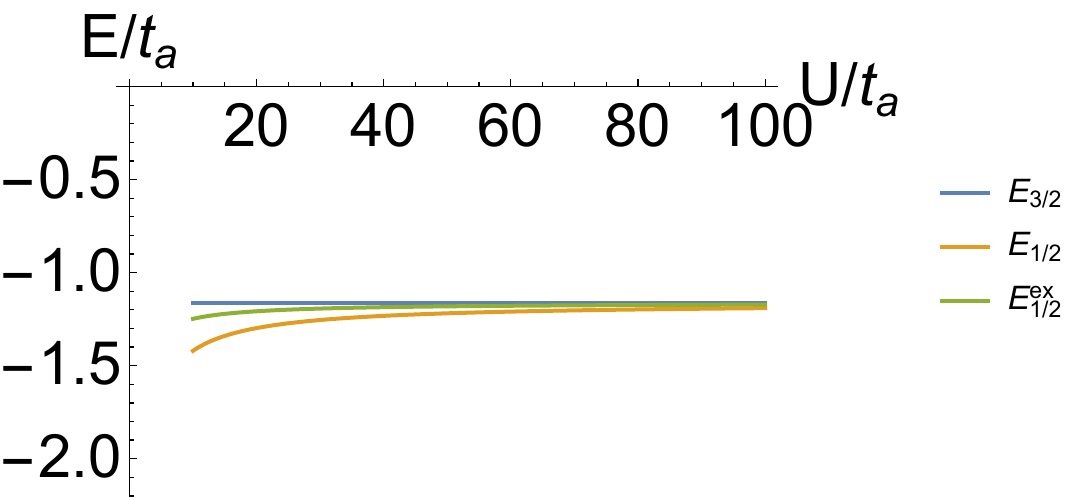}
	\caption{Plot of $E_{3/2}$, $E_{1/2}$ and $E_{1/2}^{\text{ex}}$ versus $U$ for three electrons in a four-dot linear array for $V=0$ (Top) and  $V=5t_a$ (Bottom). Here $W=V/4$.}
\end{figure}

We repeat the procedure discussed above to calculate the finite $U$ corrections to $E_{1/2}^\pm$. Then $-T^\dagger\Lambda^{-1}T$ is given by:
\begin{align}
&-T^\dagger\Lambda^{-1}T\nonumber\\
&=\bigg[\frac{-B^2t_a^2}{U}-\frac{((A+B)^2+A^2)t_a^2}{U+2W}-\frac{2A^2t_a^2}{U+2V}\bigg]\begin{pmatrix}2&1\\1&2\end{pmatrix}\nonumber\\
\end{align}

The off-diagonal terms break the $\ket{\Psi_{1/2}^+},\ket{\Psi_{1/2}^-}$ degeneracy, with the lower energy state given by:
\begin{equation}
\ket{\Psi_{1/2}}=\frac{1}{\sqrt{2}}\Big[\ket{\Psi_{1/2}^+}+\ket{\Psi_{1/2}^-}\Big]
\end{equation}

which corresponds to the spin configuration:
\begin{equation}
\frac{1}{\sqrt{6}}\big[-\ket{\uparrow\uparrow\downarrow}+2\ket{\uparrow\downarrow\uparrow}-\ket{\downarrow\uparrow\uparrow}\big]
\end{equation}

This spin configuration is the spin 1/2 state which maximizes overlap with the alternating spin configuration $\ket{\uparrow\downarrow\uparrow}$, and so the ground state of 3 electrons in a linear array of 4 dots is an antiferromagnet. The ground state energy is given by:
\begin{align}
E_{1/2}=&\,\frac{V-t_a-\sqrt{(V+t_a)^2+4t_a^2}}{2}\nonumber\\&-3t_a^2\bigg[\frac{B^2}{U}+\frac{((A+B)^2+A^2)}{U+2W}+\frac{2A^2}{U+2V}\bigg]
\end{align}	


\subsubsection{Y-Shaped Configuration}

We now model a Y-shaped configuration of four dots. We will let dots 2 through 4 be positioned at the corners of an equilateral triangle, and dot 1 be at the center, with hopping terms only between a corner dot and the center dot. Then $t_{ij}$ and $V_{ij}$ are given by:
\begin{equation}
t_{ij}=\begin{cases}
-t_a&\text{if }i\text{ or }j=1\\
0&\text{otherwise}
\end{cases}
\end{equation}
\begin{equation}
V_{ij}=\begin{cases}
V_a&\text{if }i\text{ or }j=1\\
V_d&\text{otherwise}
\end{cases}
\end{equation}

Then up to symmetry, the following electron configurations are possible:
\begin{align}
&(0,1,1,1)\;\text{ with energy: }\;3V_d\nonumber\\
&(1,1,1,0)\;\text{ with energy: }\;2V_a+V_d\nonumber\\
&(0,2,1,0)\;\text{ with energy: }\;U_0+2V_d\nonumber\\
&(2,1,0,0)\;\text{ with energy: }\;U_0+2V_a\nonumber\\
&(1,2,0,0)\;\text{ with energy: }\;U_0+2V_a
\label{eqn:yconfig}
\end{align}

We shift the total energy by $3V_d$, and define $U$ and $V$ as:
\begin{align}
U&\equiv U_0-V_d\nonumber\\
V&\equiv V_a-V_d
\label{eqn:uvdefy}
\end{align}

so that the energies of the electron configurations in eq. (\ref{eqn:yconfig}) become $0$, $2V$, $U$, $U+2V$, $U+2V$ respectively. Using the same methods as above, we construct the spin 3/2 Hamiltonian:
\begin{equation}
H_{3/2}=\begin{pmatrix}
2V&0&0&-t_a\\
0&2V&0&t_a\\
0&0&2V&-t_a\\
-t_a&t_a&-t_a&0
\end{pmatrix}
\end{equation}

which has a nondegenerate ground state with energy:
\begin{equation}
E_{3/2}=V-\sqrt{V^2+3t_a^2}
\end{equation}

given by:
\begin{equation}
\Psi_{3/2}=\frac{1}{\sqrt{3+\frac{9t_a^2}{E_{3/2}^2}}}\begin{pmatrix}1\\-1\\1\\3t_a/(-E_{3/2})\end{pmatrix}
\label{eqn:abydefref}
\end{equation}


The first excited state has energy $2V$, and so the spin gap is given by the difference $2V-E_{3/2}$.

For the spin 1/2 case, in the infinite $U$ limit, the Hamiltonian separates into a block-diagonal matrix with two blocks, where the basis for each block is given by:
\begin{equation}
\begin{matrix}
\ket{\psi^\pm_1\psi^\pm_2\psi^\pm_30},&\ket{\psi^\mp_1\psi^\mp_20\psi^\mp_3},&\ket{\psi^\pm_10\psi^\pm_2\psi^\pm_3},&\ket{0\psi^\pm_1\psi^\pm_2\psi^\pm_3}
\end{matrix}
\end{equation}

In this basis, the two blocks of the spin 1/2 Hamiltonian $H_{1/2}^\pm$ are given by:
\begin{equation}
H_{1/2}^\pm=\begin{pmatrix}
2V&0&0&-t_ae^{\mp\frac{2\pi i}{3}}\\
0&2V&0&t_ae^{\mp\frac{2\pi i}{3}}\\
0&0&2V&-t_a\\
-t_ae^{\pm\frac{2\pi i}{3}}&t_ae^{\pm\frac{2\pi i}{3}}&-t_a&0
\end{pmatrix}
\end{equation}

which is identical to $H_{3/2}$ up to a phase redefinition of some of the states. Therefore in the infinite $U$ limit, $E_{1/2}^\pm=E_{3/2}$, and thus for finite $U$, the system cannot exhibit ferromagnetism, since the finite $U$ corrections will lower the energy of the spin 1/2 states.

\subsubsection{Y-Shaped Configuration With N.N.N. Hopping}

We now add a next nearest neighbor hopping term $t_d$ between the outer corners of the Y-shaped configuration. Then $t_{ij}$ is given by:
\begin{equation}
t_{ij}=\begin{cases}
-t_a&\text{if }i\text{ or }j=1\\
-t_d&\text{otherwise}
\end{cases}
\end{equation}

The same electron configurations as in eq. (\ref{eqn:yconfig}) above are possible. We again shift the total energy by $3V_2$, and define $U$ and $V$ as in eq. (\ref{eqn:uvdefy}). Using the same methods as above, we construct the Hamiltonian:
\begin{equation}
H_{3/2}=\begin{pmatrix}
2V&-t_d&t_d&-t_a\\
-t_d&2V&-t_d&t_a\\
t_d&-t_d&2V&-t_a\\
-t_a&t_a&-t_a&0
\end{pmatrix}
\end{equation}

which has a nondegenerate ground state with energy:
\begin{equation}
E_{3/2}=V+t_d-\sqrt{(V+t_d)^2+3t_a^2}
\end{equation}

The first excited state has energy $2V-t_d$.

We construct the spin 1/2 Hamiltonian in the basis given by eq. (\ref{eqn:s12basis3e4d}) as follows:

\begin{widetext}
	\begin{equation}
	H_{1/2}=\begin{pmatrix}
	2V&-t_d&0&-t_ae^{\frac{-2\pi i}{3}}&0&0&t_de^{\frac{-2\pi i}{3}}&0\\
	-t_d&2V&-t_d&0&0&0&0&t_ae^{\frac{2\pi i}{3}}\\
	0&-t_d&2V&-t_a&t_de^{\frac{-2\pi i}{3}}&0&0&0\\
	-t_ae^{\frac{2\pi i}{3}}&0&-t_a&0&0&t_ae^{\frac{2\pi i}{3}}&0&0\\
	0&0&t_de^{\frac{2\pi i}{3}}&0&2V&-t_d&0&-t_ae^{\frac{2\pi i}{3}}\\
	0&0&0&t_ae^{\frac{-2\pi i}{3}}&-t_d&2V&-t_d&0\\
	t_de^{\frac{2\pi i}{3}}&0&0&0&0&-t_d&2V&-t_a\\
	0&t_ae^{\frac{-2\pi i}{3}}&0&0&-t_ae^{\frac{-2\pi i}{3}}&0&-t_a&0
	\end{pmatrix}
	\end{equation}
\end{widetext}

This matrix has two degenerate ground states with energy given by the smallest root of a cubic polynomial $P(E_{1/2})=0$, where $P(E)$ is given by:
\begin{align}
P(E)=E^3-4VE^2+(-3t_a^2-t_d^2+4V^2)E+6t_a^2V\nonumber\\
\end{align}

To compare $E_{1/2}$ with $E_{3/2}$, one can show that $P(E_{3/2})>0$ for $0<t_d<t_a$. This implies that there must be a root of $P(E)$ which lies to the left of $E_{3/2}$, and thus $E_{1/2}<E_{3/2}$. Therefore the ground state is not ferromagnetic.

\subsection{Summary}

We have explored many different plaquette geometries in the presence of long-range Coulomb interactions, with and without next nearest neighbor hopping. We have found that in these systems, Nagaoka ferromagnetism is robust to the presence of long-range Coulomb interactions, and is present even if the plaquette is rectangular rather than square. We argued that next nearest neighbor hopping destroys Nagaoka ferromagnetism, and derived conditions for the value of $t_d$ where this transition occurs for both the square and rectangular geometries. for completeness, we showed that other geometries such as a linear array and Y-shaped configuration have an antiferrromagnetic ground state. We present these findings in a table below:

\begin{widetext}	
\begin{center}
\begin{tabular}{ |c|c|c|c|c|c|c| } 
	\hline
	Sec.&dot&n.n.n.&$E_{3/2}$&$E_{1/2}$&Spin&Ferro-\\
	num.&config.&hopping&&for $U\rightarrow\infty$&gap&magnetism?\\\hline
	&&&&&&\\\hline
	1&square&no&$-2t_a$&$-\sqrt{3}t_a$&$2t_a$&yes\\\hline
	2&square&yes&$-2t_a+t_d$&$-\sqrt{3t_a^2+t_d^2}$&$2t_a-2t_d$&if $t_d<t_a/4$\\\hline
	3&rectangle&no&$-t_a-t_b$&$-\sqrt{t_a^2+t_at_b+t_b^2}$&$2t_b$&yes\\\hline
	4&rectangle&yes&$-t_a-t_b+t_d$&$-\sqrt{t_a^2+t_b^2+t_d^2+t_at_b+t_at_d-t_bt_d}$&$2t_b-2t_d$&if $t_d<\frac{t_at_b}{3t_a+t_b}$\\\hline
	5&linear&no&$\frac{1}{2}\begin{pmatrix}V-t_a\qquad\qquad\\-\sqrt{(V+t_a)^2+4t_a^2}\end{pmatrix}$&$\frac{1}{2}\begin{pmatrix}V-t_a\qquad\qquad\\-\sqrt{(V+t_a)^2+4t_a^2}\end{pmatrix}$&$\Delta_{\text{lin}}$&no\\\hline
	6&Y-shaped&no&$V-\sqrt{V^2+3t_a^2}$&$V-\sqrt{V^2+3t_a^2}$&$2V-E_{3/2}$&no\\\hline
	7&Y-shaped&yes&$\begin{matrix}V+t_d\qquad\qquad\quad\\-\sqrt{(V+t_d)^2+3t_a^2}\end{matrix}$&given by $P(E_{1/2})=0$&$\begin{matrix}2V-t_d\\-E_{3/2}\end{matrix}$&no\\\hline
\end{tabular}
\\
\end{center}

$\Delta_{\text{lin}}=t_a+\frac{1}{2}\Big(\sqrt{(V+t_a)^2+4t_a^2}-\sqrt{(V-t_a)^2+4t_a^2}\Big)$

$P(E_{1/2})=E_{1/2}^3-4VE_{1/2}^2+(-3t_a^2-t_d^2+4V^2)E_{1/2}+6t_a^2V$

\begin{flalign*}
&V\equiv\begin{cases}
V_a-V_{3a} & \text{for sec. 5}\\
V_a-V_d & \text{for sec. 6 \& 7}
\end{cases}&&
\end{flalign*}

\end{widetext}

\section{Four Electrons In Four Dots}

\subsection{General Method}

We now consider a half-filled band consisting of four electrons and four dots in an arbitrary configuration, for large $U_0$. It is well-known that for large systems, the ground state of a half-filled band is antiferromagnetic; however, we show that a four dot plaquette can have a partially ferromagnetic spin-1 ground state for certain geometries.

The lowest energy states will be in the $(1,1,1,1)$ configuration, and we will shift the energy of our Hamiltonian to account for the Coulomb interaction energy in this configuration. Thus, by definition, the spin 2 states, which are not affected by tunneling, have energy $E_2=0$, and in the infinite $U$ limit, the spin 0 and 1 states have 0 energy as well. We will define $U_{i(j)}$ to be the Coulomb interaction energy of the $(2,1,1,0)$ configuration states with two electrons in dot $i$ and no electrons in dot $j$, again offset by the energy of the $(1,1,1,1)$ state. We ignore the $(2,2,0,0)$ states, as they are not connected to the $(1,1,1,1)$ states by a single tunneling operation, and thus will have no effect on the ground state energies to order $t^2/U$.

Our strategy is the same as when finding the finite $U$ corrections in the previous section. We list all low energy states with a given spin. Since these will all be in the $(1,1,1,1)$ configuration, they will be degenerate to leading order. We then list the relevant high energy states, and let $\Lambda$ be the diagonal matrix with entries given by the energies of the high energy states, and let the entries of $T$ be given by the matrix elements of $H$ between a low and a high energy state. Then the first order corrections in $t^2/U$ to the energies of the low energy states are given by diagonalizing the matrix $-T^\dagger\Lambda^{-1} T$. This will potentially also break the degeneracy, as long as $-T^\dagger\Lambda^{-1} T$ is not proportional to the identity matrix.

\subsubsection{Spin 0 States}

There are two states with total spin 0 for electrons in the $(1,1,1,1)$ configuration:
\begin{align}
\ket{\Psi_0^\pm}=\frac{1}{\sqrt{6}}\bigg[&\,e^{\pm\frac{2\pi i}{3}}\ket{\uparrow\uparrow\downarrow\downarrow}+\ket{\uparrow\downarrow\uparrow\downarrow}+e^{\mp\frac{2\pi i}{3}}\ket{\uparrow\downarrow\downarrow\uparrow}\nonumber\\&+e^{\mp\frac{2\pi i}{3}}\ket{\downarrow\uparrow\uparrow\downarrow}+\ket{\downarrow\uparrow\downarrow\uparrow}+e^{\pm\frac{2\pi i}{3}}\ket{\downarrow\downarrow\uparrow\uparrow}\bigg]
\end{align}

There are 24 high energy states connected to $\ket{\Psi_0^\pm}$ by a single tunneling operation, corresponding to all permutations of $\ket{\,\ud\uparrow\,\downarrow0}$. However, due to conservation of spin, only states where the two single electrons form a spin singlet will contribute, and thus we need only consider 12 states. We calculating matrix elements between these states and $\ket{\Psi_0^\pm}$, we obtain the matrix $-T^\dagger\Lambda^{-1}T$:
\begin{equation}
-T^\dagger\Lambda^{-1}T=-\sum_{i\ne j}\frac{t_{ij}^2}{U_{i(j)}}\begin{pmatrix}
1&e^{-i\varphi_{ij}}\\e^{i\varphi_{ij}}&1\end{pmatrix}
\end{equation}

where $\varphi_{ij}$ is given by:
\begin{equation}
\varphi_{ij}=\begin{cases}
\frac{\pi}{3}&\text{if }\{i,j\}=\{1,2\}\text{ or }\{3,4\}\\
\pi&\text{if }\{i,j\}=\{1,3\}\text{ or }\{2,4\}\\
\frac{5\pi}{3}&\text{if }\{i,j\}=\{1,4\}\text{ or }\{2,3\}
\end{cases}
\end{equation}

Thus, to order $t^2/U$, the total energy of the spin 0 ground state is:
\begin{equation}
E_0=-\sum_{i\ne j}\frac{t_{ij}^2}{U_{i(j)}}-\Bigg|\sum_{i\ne j}\frac{t_{ij}^2}{U_{i(j)}}e^{i\varphi_{ij}}\Bigg|
\end{equation}

\subsubsection{Spin 1 States}

To investigate the spin 1 states, we consider the subspace where $S_z=1$. There are three states with total spin 1 for electrons in the $(1,1,1,1)$ configuration:
\begin{align}
\ket{\Psi_1^1}&=\frac{1}{2}\Big[\ket{\uparrow\uparrow\uparrow\downarrow}+\ket{\uparrow\uparrow\downarrow\uparrow}-\ket{\uparrow\downarrow\uparrow\uparrow}-\ket{\downarrow\uparrow\uparrow\uparrow}\Big]\nonumber\\
\ket{\Psi_1^2}&=\frac{1}{2}\Big[\ket{\uparrow\uparrow\uparrow\downarrow}-\ket{\uparrow\uparrow\downarrow\uparrow}+\ket{\uparrow\downarrow\uparrow\uparrow}-\ket{\downarrow\uparrow\uparrow\uparrow}\Big]\nonumber\\
\ket{\Psi_1^3}&=\frac{1}{2}\Big[\ket{\uparrow\uparrow\uparrow\downarrow}-\ket{\uparrow\uparrow\downarrow\uparrow}-\ket{\uparrow\downarrow\uparrow\uparrow}+\ket{\downarrow\uparrow\uparrow\uparrow}\Big]
\label{eqn:4e4ds1states}
\end{align}

There are 12 high energy states connected to $\ket{\Psi_1^i}$, given by all permutations of $\ket{\ud\uparrow\uparrow 0}$. Calculating matrix elements between these states and $\ket{\Psi_1^i}$, we find that $-T^\dagger\Lambda^{-1}T$ is given by:
\begin{align}
-T^\dagger\Lambda^{-1}T&=-\sum_{i\ne j}\frac{t_{ij}^2}{U_{i(j)}}\mathbb{1}+\nonumber\\&\begin{pmatrix}
A_{12}+A_{34}&A_{23}-A_{14}&A_{13}-A_{24}\\A_{23}-A_{14}&A_{13}+A_{24}&A_{12}-A_{34}\\A_{13}-A_{24}&A_{12}-A_{34}&A_{14}+A_{23}
\end{pmatrix}
\end{align}

where $A_{ij}$ is given by:
\begin{equation}
A_{ij}=t_{ij}^2\bigg(\frac{1}{U_{i(j)}}+\frac{1}{U_{j(i)}}\bigg)
\end{equation}

\subsection{Ground State Calculations}

\subsubsection{Square with no Diagonal Hopping}

For four dots in a square, with no diagonal hopping, we have for spin 0,
\begin{align}
(-T^\dagger\Lambda^{-1}T)_0=-\frac{t_a^2}{U}\begin{pmatrix}8&4\\4&8\end{pmatrix}
\end{align}

where $U\equiv U_0-V_a$. The off-diagonal terms break the degeneracy, and the ground state and energy is given by:

\begin{align}
\ket{\Psi_0}&=\frac{1}{2\sqrt{3}}\bigg[-\ket{\uparrow\uparrow\downarrow\downarrow}+2\ket{\uparrow\downarrow\uparrow\downarrow}-\ket{\uparrow\downarrow\downarrow\uparrow}\nonumber\\&\qquad\qquad-\ket{\downarrow\uparrow\uparrow\downarrow}+2\ket{\downarrow\uparrow\downarrow\uparrow}-\ket{\downarrow\downarrow\uparrow\uparrow}\bigg]\\
E_0&=-12\frac{t_a^2}{U}
\end{align}

We note that as expected, this is the spin 0 state which maximizes overlap with the antiferromagnetic configurations $\ket{\uparrow\downarrow\uparrow\downarrow}$ and $\ket{\downarrow\uparrow\downarrow\uparrow}$. For spin 1, we have
\begin{align}
&(-T^\dagger\Lambda^{-1}T)_1=-\frac{t_a^2}{U}\begin{pmatrix}4&0&0\\0&8&0\\0&0&4\end{pmatrix}\\
&E_1=-8\frac{t_a^2}{U}
\end{align}

Here the degeneracy is also broken, and the ground state is given by $\ket{\Psi_1^2}$ as defined in eq. (\ref{eqn:4e4ds1states}).

\subsubsection{Square With Diagonal Hopping}

For four dots in a square, with diagonal hopping, we have
\begin{align}
&(-T^\dagger\Lambda^{-1}T)_0=-\frac{t_a^2}{U}\begin{pmatrix}8&4\\4&8\end{pmatrix}-\frac{t_d^2}{U+V}\begin{pmatrix}4&-4\\-4&4\end{pmatrix}\\
&E_0=-12\frac{t_a^2}{U}
\end{align}

where $U\equiv U_0-V_a$ and $V\equiv V_a-V_d$. For spin 1,
\begin{align}
&(-T^\dagger\Lambda^{-1}T)_1=-\frac{t_a^2}{U}\begin{pmatrix}4&0&0\\0&8&0\\0&0&4\end{pmatrix}-\frac{t_d^2}{U+V}\begin{pmatrix}4&0&0\\0&0&0\\0&0&4\end{pmatrix}\\
&E_1=-8\frac{t_a^2}{U}
\end{align}

Interestingly, diagonal hopping for a square does not affect the ground state energies $E_0$ or $E_1$, and only serves to decrease the energy of the excited states. This can be understood by noticing that in each of the ground states, spins at opposite corners of the square (dots 1 and 3 or dots 2 and 4) only occur in a triplet configuration. This is necessary to allow adjacent spins to anti-align as much as possible.

\subsubsection{Rectangle}

For four dots in a rectangle, with no diagonal hopping, we have
\begin{align}
&(-T^\dagger\Lambda^{-1}T)_0=\nonumber\\&-4\begin{pmatrix}\frac{t_a^2}{U}+\frac{t_b^2}{U+V}&\frac{t_a^2}{U}e^{\frac{-\pi i}{3}}+\frac{t_b^2}{U+V}e^{\frac{\pi i}{3}}\\\frac{t_a^2}{U}e^{\frac{\pi i}{3}}+\frac{t_b^2}{U+V}e^{\frac{-\pi i}{3}}&\frac{t_a^2}{U}+\frac{t_b^2}{U+V}\end{pmatrix}\\
&E_0=-4\Bigg[\frac{t_a^2}{U}+\frac{t_b^2}{U\!+\!V}+\sqrt{\frac{t_a^4}{U^2}+\frac{t_b^4}{(U\!+\!V)^2}-\frac{t_a^2t_b^2}{U(U\!+\!V)}}\Bigg]
\end{align}

where $U\equiv U_0-V_a$, and $V\equiv V_a-V_b$. For spin 1,
\begin{align}
&(-T^\dagger\Lambda^{-1}T)_1=-4\bigg[\frac{t_a^2}{U}+\frac{t_b^2}{U+V}\bigg]\mathbb{1}+4\begin{pmatrix}\frac{t_a^2}{U}&0&0\\0&0&0\\0&0&\frac{t_b^2}{U+V}\end{pmatrix}\\
&E_1=-4\Bigg[\frac{t_a^2}{U}+\frac{t_b^2}{U+V}\Bigg]
\end{align}

We note that the spin 1 ground state remains the same as in the square case, while the spin 0 ground state rotates, essentially in such a way as to include a greater weight to singlets across the shorter edge of the rectangle than the longer edge. This must be the case, as when $t_b\rightarrow 0$, the ground state must become two spin singlets.

\subsubsection{Rectangle With Diagonal Hopping}

For four dots in a rectangle, with diagonal hopping, we have
\begin{align}
&(-T^\dagger\Lambda^{-1}T)_0=\nonumber\\&-4\begin{pmatrix}\quad\frac{t_a^2}{U}\!+\!\frac{t_b^2}{U+V}\!+\!\frac{t_d^2}{U+W}\qquad\frac{t_a^2}{U}e^{\frac{-\pi i}{3}}\!+\!\frac{t_b^2}{U+V}e^{\frac{\pi i}{3}}\!-\!\frac{t_d^2}{U+W}\\\frac{t_a^2}{U}e^{\frac{\pi i}{3}}\!+\!\frac{t_b^2}{U+V}e^{\frac{-\pi i}{3}}\!-\!\frac{t_d^2}{U+W}\qquad\frac{t_a^2}{U}\!+\!\frac{t_b^2}{U+V}\!+\!\frac{t_d^2}{U+W}\quad\end{pmatrix}\\
&E_0=-4\Bigg[\frac{t_a^2}{U}+\frac{t_b^2}{U\!+\!V}+\frac{t_d^2}{U\!+\!W}\nonumber\\&\qquad+\bigg(\frac{t_a^4}{U^2}+\frac{t_b^4}{(U\!+\!V)^2}+\frac{t_d^4}{(U\!+\!W)^2}\nonumber\\&\qquad-\frac{t_a^2t_b^2}{U(U\!+\!V)}-\frac{t_a^2t_d^2}{U(U\!+\!W)}-\frac{t_b^2t_d^2}{(U\!+\!V)(U\!+\!W)}\bigg)^{1/2}\Bigg]
\end{align}

where $U\equiv U_0-V_a$, $V\equiv V_a-V_b$, and $W=V_a-V_d$. For spin 1,
\begin{align}
&(-T^\dagger\Lambda^{-1}T)_1=\nonumber\\&-4\bigg[\frac{t_a^2}{U}+\frac{t_b^2}{U+V}+\frac{t_d^2}{U+W}\bigg]\mathbb{1}+4\begin{pmatrix}\frac{t_a^2}{U}&0&0\\0&\frac{t_d^2}{U+W}&0\\0&0&\frac{t_b^2}{U+V}\end{pmatrix}\\
&E_1=-4\Bigg[\frac{t_a^2}{U}+\frac{t_b^2}{U+V}\Bigg]
\end{align}

Again the the spin 1 ground state is unaffected by the presence of diagonal hopping, for the same reason discussed above. However, the diagonal hopping terms do affect the spin 0 state, since the imbalance between $t_a$ and $t_b$ causes opposite spins to no longer only appear in triplets.

\subsubsection{Linear Array}

For four dots in a line, we have
\begin{align}
&(-T^\dagger\Lambda^{-1}T)_0=\nonumber\\&-2t_a^2\begin{pmatrix}\frac{1}{U}\!+\!\frac{1}{U+2V}\!+\!\frac{1}{U+V}&(\frac{1}{U}\!+\!\frac{1}{U+2V})e^{\frac{-\pi i}{3}}\!+\!\frac{e^{\frac{\pi i}{3}}}{U+V}\\(\frac{1}{U}\!+\!\frac{1}{U+2V})e^{\frac{\pi i}{3}}\!+\!\frac{e^{\frac{-\pi i}{3}}}{U+V}&\frac{1}{U}\!+\!\frac{1}{U+2V}\!+\!\frac{1}{U+V}\end{pmatrix}\\
&E_0=-2t_a^2\Bigg[\frac{1}{U}+\frac{1}{U\!+\!2V}+\frac{1}{U\!+\!V}\nonumber\\&+\sqrt{\big(\frac{1}{U}\!+\!\frac{1}{U\!+\!2V}\big)^2+\frac{1}{(U\!+\!V)^2}-\frac{1}{U\!+\!V}\big(\frac{1}{U}\!+\!\frac{1}{U\!+\!2V}\big)}\Bigg]
\end{align}

where $U\equiv U_0-2V_a+V_{3a}$, and $V\equiv V_a-V_{3a}$. For spin 1,
\begin{align}
&(-T^\dagger\Lambda^{-1}T)_1=-2t_a^2\Big[\frac{1}{U}+\frac{1}{U+2V}+\frac{1}{U+V}\Big]\mathbb{1}\nonumber\\&\qquad\qquad\qquad+2t_a^2\begin{pmatrix}\frac{1}{U}+\frac{1}{U+2V}&\frac{1}{U+V}&0\\\frac{1}{U+V}&0&0\\0&0&\frac{1}{U+V}\end{pmatrix}\\
&E_1=\frac{-t_a^2}{U}-\frac{t_a^2}{U+2V}-\frac{2t_a^2}{U+V}\nonumber\\&\qquad-t_a^2\sqrt{\big(\frac{1}{U}+\frac{1}{U+2V}\big)^2+\frac{4}{(U+V)^2}}
\end{align}

In the limit where $V\rightarrow0$, this reduces to $E_0=-(6+2\sqrt{3})t_a^2/U$ and $E_1=-(4+2\sqrt{2})t_a^2/U$.

\subsubsection{Y-Shaped Configuration}

For four dots in a Y-shaped configuration, we have
\begin{align}
&(-T^\dagger\Lambda^{-1}T)_0=-\Big(\frac{t_a^2}{U}+\frac{t_a^2}{U+4V}\Big)\begin{pmatrix}3&0\\0&3\end{pmatrix}\\
&E_0=-3\Big(\frac{t_a^2}{U}+\frac{t_a^2}{U+4V}\Big)
\end{align}

where $U\equiv U_0-3V_a+2V_d$ and $V\equiv V_a-V_d$. Thus, the $\ket{\Psi_0^\pm}$ degeneracy remains unbroken, due to the three-fold rotational symmetry of the system. For spin 1,
\begin{align}
&(-T^\dagger\Lambda^{-1}T)_1=\Big(\frac{t_a^2}{U}+\frac{t_a^2}{U+4V}\Big)\begin{pmatrix}-2&-1&1\\-1&-2&1\\1&1&-2\end{pmatrix}\\
&E_1=-4\Big(\frac{t_a^2}{U}+\frac{t_a^2}{U+4V}\Big)
\end{align}

Interestingly, the ground state is the spin 1 state rather than the spin 0 state. This state is given by:
\begin{equation}
\ket{\Psi_1}=\frac{1}{2\sqrt{3}}\Big[\ket{\uparrow\uparrow\uparrow\downarrow}+\ket{\uparrow\uparrow\downarrow\uparrow}+\ket{\uparrow\downarrow\uparrow\uparrow}-3\ket{\downarrow\uparrow\uparrow\uparrow}\Big]
\end{equation}

which is the state maximizes the weight of the spin configuration where the center electron has opposite spin as the three corner electrons. Thus, the ground state can be thought of as antiferromagnetic in the sense that adjacent spins are anti-aligned; however, since there is an imbalance in the number of sites in the odd and even sublattices, assigning alternating spins to these sites causes a total spin of 1 rather than 0.

\subsubsection{Y-Shaped Configuration with N.N.N. Hopping}

For four dots in a Y-shaped configuration, with next nearest neighbor hopping (that is hopping between the outer corners), we have
\begin{align}
&(-T^\dagger\Lambda^{-1}T)_0=-\Big(\frac{t_a^2}{U}+\frac{t_a^2}{U+4V}+\frac{2t_d^2}{U+3V}\Big)\begin{pmatrix}3&0\\0&3\end{pmatrix}\\
&E_0=-3\Big(\frac{t_a^2}{U}+\frac{t_a^2}{U+4V}+\frac{2t_d^2}{U+3V}\Big)
\end{align}

where $U\equiv U_0-3V_a+2V_d$ and $V\equiv V_a-V_d$. For spin 1,
\begin{align}
&(-T^\dagger\Lambda^{-1}T)_1=-2\Big(\frac{t_a^2}{U}+\frac{t_a^2}{U+4V}+\frac{2t_d^2}{U+3V}\Big)\mathbb{1}\nonumber\\&+\Big(\frac{t_a^2}{U}+\frac{t_a^2}{U+4V}-\frac{2t_d^2}{U+3V}\Big)\begin{pmatrix}0&-1&1\\-1&0&1\\1&1&0\end{pmatrix}\\
&E_1=-4\Big(\frac{t_a^2}{U}+\frac{t_a^2}{U+4V}\Big)
\end{align}

Here the presence of next nearest neighbor hopping reduces the energy of the spin 0 states, while still maintaining the $\ket{\Psi_0^\pm}$ degeneracy, as the three-fold symmetry of the system remains unbroken. The next nearest neighbor hopping terms do not affect the spin 1 ground state, however, as the spins in any of the two corners only appear in triplet configurations. Thus, as $t_d$ is increased there exists a crossover point between $E_0$ and $E_1$. $E_1<E_0$ as long as $\frac{6t_d^2}{U+3V}<\frac{t_a^2}{U}+\frac{t_a^2}{U+4V}$.

\subsection{Summary}

We have calculated the energies of the lowest energy spin 0 and spin 1 state for half-filled band in several different four-dot geometries to first order in $t^2/U$. In each case, the ground state is antiferromagnetic; however, with the Y-shaped configuration, alternating spins on each cite causes a total spin of 1 rather than 0, because there are 3 corner dots and only 1 center dot. Adding next nearest neighbor interactions reduces the energy difference between the two states, up to a critical strength at which point the spin 0 state becomes the ground state. We summarize these findings in a table:

\begin{widetext}

\begin{center}
	\begin{tabular}{ |c|c|c|c|c|c|c| } 
		\hline
		Sec.&dot&nnn.&$E_0$&$E_1$&$\;E_2\;$&$E_1<E_0$?\\
		num.&config.&hop.&to order $t^2/U$&to order $t^2/U$&&\\\hline
		&&&&&&\\\hline
		1&square&no&$-12t_a^2/U$&$-8t_a^2/U$&$0$&no\\\hline
		2&square&yes&$-12t_a^2/U$&$-8t_a^2/U$&$0$&no\\\hline
		3&rectangle&no&$\begin{matrix}\frac{-4t_a^2}{U}-\frac{4t_b^2}{U+V}\qquad\qquad\qquad\\-4\sqrt{\frac{t_a^4}{U^2}+\frac{t_b^4}{(U+V)^2}-\frac{t_a^2t_b^2}{U(U+V)}}\end{matrix}$&$\frac{-4t_a^2}{U}-\frac{4t_b^2}{U+V}$&$0$&no\\\hline
		4&rectangle&yes&$\begin{matrix}\frac{-4t_a^2}{U}-\frac{4t_b^2}{U+V}-\frac{4t_d^2}{U+W}\qquad\qquad\qquad\qquad\\-4\begin{pmatrix}\frac{t_a^4}{U^2}+\frac{t_b^4}{(U+V)^2}+\frac{t_d^4}{(U+W)^2}\qquad\qquad\\-\frac{t_a^2t_b^2}{U(U+V)}-\frac{t_a^2t_d^2}{U(U+W)}-\frac{t_b^2t_d^2}{(U+V)(U+W)}\end{pmatrix}^{\frac{1}{2}}\end{matrix}$&$\frac{-4t_a^2}{U}-\frac{4t_b^2}{U+V}$&$0$&no\\\hline
		5&linear&no&$\begin{matrix}\frac{-2t_a^2}{U}-\frac{2t_a^2}{U+2V}-\frac{2t_a^2}{U+V}\qquad\qquad\qquad\\-2t_a^2\begin{pmatrix}\big(\frac{1}{U}+\frac{1}{U+2V}\big)^2+\frac{1}{(U+V)^2}\\-\frac{1}{U+V}\big(\frac{1}{U}+\frac{1}{U+2V}\big)\quad\end{pmatrix}^{1/2}\end{matrix}$&$\begin{matrix}\frac{-t_a^2}{U}-\frac{t_a^2}{U+2V}-\frac{2t_a^2}{U+V}\qquad\qquad\\-t_a^2\sqrt{\big(\frac{1}{U}+\frac{1}{U+2V}\big)^2+\frac{4}{(U+V)^2}}\end{matrix}$&$0$&no\\\hline
		6&Y-shaped&no&$-3\Big(\frac{t_a^2}{U}+\frac{t_a^2}{U+4V}\Big)$&$-4\Big(\frac{t_a^2}{U}+\frac{t_a^2}{U+4V}\Big)$&$0$&yes\\\hline
		7&Y-shaped&yes&$-3\Big(\frac{t_a^2}{U}+\frac{t_a^2}{U+4V}+\frac{2t_d^2}{U+3V}\Big)$&$-4\Big(\frac{t_a^2}{U}+\frac{t_a^2}{U+4V}\Big)$&$0$&$\begin{matrix}\text{if }\frac{6t_d^2}{U+3V}<\\\frac{t_a^2}{U}+\frac{t_a^2}{U+4V}\end{matrix}$\\\hline
	\end{tabular}
\end{center}

\end{widetext}

Note we define $U$ and $V$ slightly differently each time:
\begin{flalign*}
&U\equiv\begin{cases}
U_0-V_a & \text{for sec. 1, 2, 3 \& 4}\\
U_0-2V_a+V_{3a} & \text{for sec. 5}\\
U_0-3V_a+2V_d & \text{for sec. 6 \& 7}\\
\end{cases}&&
\end{flalign*}

\begin{flalign*}
&V\equiv\begin{cases}
V_a-V_b & \text{for sec. 3 \& 4}\\
V_a-V_{3a} & \text{for sec. 5}\\
V_a-V_d & \text{for sec. 6 \& 7}\\
\end{cases}&&\\
\\
&W\equiv V_a-V_d&&
\end{flalign*}

\section{Four Electrons in Five Dots}

\subsection{Model}

\begin{figure}[!htb]
	\includegraphics[width=.5\columnwidth]{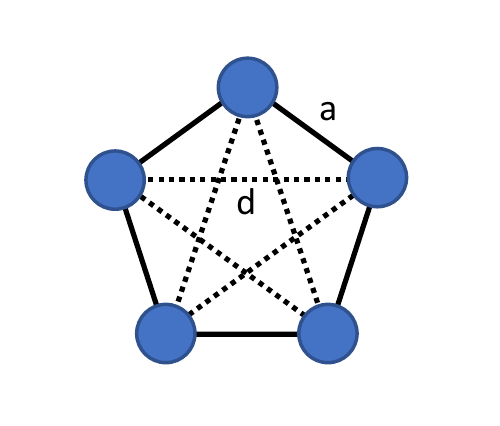}
	\caption{A depiction of a ring of 5 dots. Solid lines depict nearest-neighbor hopping terms and Coulomb interactions, and dashed lines long-range Coulomb interactions.}
	\label{fig:ring5}
\end{figure}

We now consider a ring of five dots with four electrons. This does not satisfy the Nagaoka condition, and thus we do not predict the ground state to be ferromagnetic. The Hamiltonian is given by eq. (\ref{eqn:hgen}), with $t_{ij}$ and $V_{ij}$ given as follows:
\begin{equation}
t_{ij}=\begin{cases}
-t_a&\text{if }i-j=\pm1\mod 5\\
0&\text{otherwise}
\end{cases}
\end{equation}
\begin{equation}
V_{ij}=\begin{cases}
V_a&\text{if }i-j=\pm1\mod 5\\
V_d&\text{if }i-j=\pm2\mod 5
\end{cases}
\end{equation}

Up to symmetry, only one low energy electron configuration is possible. There are also three high energy configurations that are connected to the low energy states by a single tunneling operation. These are:
\begin{align}
&(1,1,1,1,0)\;\text{ with energy: }\;3V_a+3V_d\nonumber\\
&(2,0,1,1,0)\;\text{ with energy: }\;U_0+V_a+4V_d\nonumber\\
&(2,1,0,1,0)\;\text{ with energy: }\;U_0+2V_a+3V_d\nonumber\\
&(2,1,1,0,0)\;\text{ with energy: }\;U_0+3V_a+2V_d
\label{eqn:5config}
\end{align}
We shift the total energy of the Hamiltonian by $3V_a+3V_d$, and define $U$ and $V$ as:
\begin{align}
&U\equiv U_0-2V_a+V_d\nonumber\\
&V\equiv V_a-V_d
\end{align}
so that the energies of the electrons configurations in eq. (\ref{eqn:5config}) become $0$, $U$, $U+V$, and $U+2V$ respectively.

\subsection{Ground State Calculation}

\subsubsection{Spin 2}

We proceed in a similar fashion as above. For spin 2, there are five states for each value of $S_z$ corresponding to the position of the hole, since there is only one spin configuration for a given value of $S_z$ that has spin 2. For $S_z=2$, these states are:
\begin{equation}
\begin{matrix}
\ket{\uparrow\,\uparrow\,\uparrow\,\uparrow 0},&\ket{\uparrow\,\uparrow\,\uparrow 0\uparrow},&\ket{\uparrow\,\uparrow 0\uparrow\,\uparrow},&\ket{\uparrow 0\uparrow\,\uparrow\,\uparrow},&\ket{0\uparrow\,\uparrow\,\uparrow\,\uparrow}
\end{matrix}
\label{eqn:basis2}
\end{equation}

In this basis, the spin 2 Hamiltonian is given as follows:
\begin{equation}
H_2=-t_a\begin{pmatrix}
0&1&0&0&-1\\
1&0&1&0&0\\
0&1&0&1&0\\
0&0&1&0&1\\
-1&0&0&1&0
\end{pmatrix}
\end{equation}

Here the sign in the (1,5) elements is due to Fermi exchange statistics. There are two degenerate ground states to this Hamiltonian given by:
\begin{equation}
\ket{\Psi_2^\pm}=\frac{1}{\sqrt{5}}\begin{pmatrix}
1&e^{\pm\frac{\pi i}{5}}&e^{\pm\frac{2\pi i}{5}}&e^{\pm\frac{3\pi i}{5}}&e^{\pm\frac{4\pi i}{5}}
\end{pmatrix}^T
\end{equation}

with energy:
\begin{equation}
E_2=-\frac{1+\sqrt{5}}{2}t_a
\end{equation} 

\subsubsection{Spin 1}

We now consider the spin 1 subspace. We define the following spin configurations:
\begin{align}
\ket{\psi_1^j}=\frac{1}{2}\Big[&\,\ket{\uparrow\uparrow\uparrow\downarrow}+e^{j\frac{\pi i}{2}}\ket{\uparrow\uparrow\downarrow\uparrow}\nonumber\\&+e^{2j\frac{\pi i}{2}}\ket{\uparrow\downarrow\uparrow\uparrow}+e^{3j\frac{\pi i}{2}}\ket{\downarrow\uparrow\uparrow\uparrow}\Big]
\end{align}
for $j$ between 1 and 3. We see that cycling the spins will return the same state with an extra phase $e^{j\frac{\pi i}{2}}$. The orbital part will be similar to the spin 2 case discussed above, and thus the spin 1 Hamiltonian will be given by a block-diagonal matrix, with blocks given as follows:

\begin{equation}
H_1^j=-t_a\begin{pmatrix}
0&1&0&0&-e^{j\frac{\pi i}{2}}\\
1&0&1&0&0\\
0&1&0&1&0\\
0&0&1&0&1\\
-e^{-j\frac{\pi i}{2}}&0&0&1&0
\end{pmatrix}
\end{equation}

This has a nondegenerate ground state with energy $E_1=-2t_a$. The ground state has spin configuration given by $\ket{\psi_1^2}$, and orbital part $\frac{1}{\sqrt{5}}(1\;1\;1\;1\;1)^T$.

\subsubsection{Spin 0}

Finally, we examine the spin 0 subspace. There are two spin configurations, which we define as follows:

\begin{align}
\ket{\psi_0^0}&=\frac{1}{2\sqrt{3}}\bigg[-\ket{\uparrow\uparrow\downarrow\downarrow}+2\ket{\uparrow\downarrow\uparrow\downarrow}-\ket{\uparrow\downarrow\downarrow\uparrow}\nonumber\\&\qquad\qquad-\ket{\downarrow\uparrow\uparrow\downarrow}+2\ket{\downarrow\uparrow\downarrow\uparrow}-\ket{\downarrow\downarrow\uparrow\uparrow}\bigg]\\
\ket{\psi_0^1}&=\frac{1}{2}\bigg[\ket{\uparrow\uparrow\downarrow\downarrow}-\ket{\uparrow\downarrow\downarrow\uparrow}-\ket{\downarrow\uparrow\uparrow\downarrow}+\ket{\downarrow\downarrow\uparrow\uparrow}\bigg]
\end{align}

We note that cycling the spins of $\ket{\psi_0^j}$ returns the same state with an additional phase $(-1)^j\ket{\psi_0^j}$. The the spin 0 Hamiltonian will be a block-diagonal matrix with blocks:

\begin{equation}
H_0^j=-t_a\begin{pmatrix}
0&1&0&0&(-1)^{j+1}\\
1&0&1&0&0\\
0&1&0&1&0\\
0&0&1&0&1\\
(-1)^{j+1}&0&0&1&0
\end{pmatrix}
\end{equation}

This also has a nondegenerate ground state with energy $E_0=-2t_a$. This state has spin configuration given by $\ket{\psi_0^1}$, and orbital part $\frac{1}{\sqrt{5}}(1\;1\;1\;1\;1)^T$.

\subsubsection{Finite $U$ Corrections}

As before, the spin 2 energy is exact for finite $U$, since the Pauli exclusion principle forbids any other states than the five examined. Additionally, since neither the spin 1 nor spin 0 ground states are degenerate with other states of the same spin, we simply use nondegenerate perturbation theory to calculate the leading order correction to the energy. We find that to order $t^2/U$, the energy of the lowest energy spin 1 state is given by:
\begin{equation}
E_1=-2t_a-4\frac{t_a^2}{U}-2\frac{t_a^2}{U+V}-2\frac{t_a^2}{U+2V}+O\Big(\frac{t_a^3}{U^2}\Big)
\end{equation}
and the energy of the lowest energy spin 0 state is given by:
\begin{equation}
E_0=-2t_a-2\frac{t_a^2}{U}-\frac{t_a^2}{U+V}-\frac{t_a^2}{U+2V}+O\Big(\frac{t_a^3}{U^2}\Big)
\end{equation}

Thus, for finite $U$, the ground state of the system is the spin 1 state. This means the ground state is partially ferromagnetic.

\section{Conclusion}

We have theoretically considered 4-dot quantum arrays in several different geometries investigating analytically within a simple, but semi-realistic, model the existence or not of Nagaoka-type ferromagnetic ground states.  Our work includes distant-neighbor hopping and distant-neighbor Coulomb coupling within a one orbital (with two spins) per dot model.  Although the interaction is always finite in our system we find several situations where Nagaoka-type ferromagnetism should emerge provided the kinetic and potential energies obey certain constraints (which we derive).  We calculate the spin gap for our system, and obtain the difference in energies between the ferromagnetic ground state and other nearby ground states.  We also provide results for a 5-dot ring with 4 electrons, finding a partially ferromagnetic ground state.  We believe that our predictions are experimentally testable in currently available quantum dot arrays as long as there is sufficient control over the system (i.e. hopping matrix elements, number of electrons in the system) and the temperature is low.  In principle, one can try to numerically calculate the hopping and the interaction matrix elements for a given system of coupled dots to make the prediction quantitative.  We, however, do not believe that such an endeavor, which would be numerically very demanding involving large configuration interaction calculations\cite{HuPRA2000,HuPRA2001,NielsenPRB2013} for the coupled dot system, is particularly useful since the necessary information for the quantum confinement in each dot is unknown and therefore, the results would be numerically unreliable.  Since all the matrix elements of hopping and interaction entering the model are likely to be exponentially sensitive to the unknown dot confinement potential, our phenomenological approach using model parameters based on a delta function confinement model is likely to have reasonable qualitative accuracy.  In particular, our specific predictions on which geometry would lead to ferromagnetism and which would not and the conditions necessary for obtaining full or partial ferromagetism in the ground states of different arrays should motivate experiments in current semiconductor dot based qubit structures where the observation of different types of nontrivial magnetic ground states could be construed as quantum emulation of interacting Hamiltonians in small systems.  We think that the experimental control already achieved in the laboratory for semiconductor qubit systems should enable the community to see various magnetic ground states in quantum dot plaquettes as predicted in our theory.

\acknowledgements

This work is supported by the Laboratory for Physical Sciences.

\end{document}